# Intra-band and Inter-band Plasmons in Transition Metal Dichalcogenide-Graphene van der Waals Heterostructures

**Partha Goswami**

D.B.College, University of Delhi, Kalkaji, New Delhi-110019, India
physicsgoswami@gmail.com

**Abstract**. In an earlier work, it was shown that the dispersion of the van der Waals heterostructures (vdWHs) of graphene monolayer on 2D transition metal dichalcogenide (GrTMD) substrate comprises of the spin-split, gapped bands. Using these it was also shown that the intra-band plasmon dispersion for the finite doping and the long wavelength limit involves the $q^{2/3}$ (unconventional) behaviour and not the well known $q^{1/2}$ behaviour. In this paper the optical conductivity and the optical properties of this vdWH are reported. We find that there are no (inter-band) plasmons for electron doping as long as the Gr-TMD hetero-structure is surrounded by the positive dielectric constant materials above and below. However, if the heterostructure is surrounded by negative dielectric constant (NDC) materials, one finds that there would be acoustic plasmons. For the hole doping case NDC material is not required. We find that the gate voltage tunable intra-band acoustic plasmons also emerge when the carrier density is greater than a moderately critical value. The stronger incarceration capability of GrTMD plasmon compared to that of doped graphene is another notable outcome of the present work. One finds that the intra-band absorbance is decreases with the frequency at a given gate voltage and increases with the gate voltage at a given frequency; the inter-band transmittance, however, is a decreasing function of frequency and the gate voltage.

## 1. Introduction

In an earlier work **[1]**, here-in-after referred to as I, it was shown that the dispersion of the Van der Waals heterostructures (vdWHs) of graphene monolayer on 2D transition metal dichalcogenide (Gr-TMD) substrate comprises of the spin-split, gapped bands. The study of the interesting and functional possibilities related to the optical conductivity of the present vdWH is reported here. One of the major effects of the substrate engineered enhanced spin-orbit coupling (SOC) is to create a bulk band gap together with the 'avoided crossing' between conduction and valence bands with opposite spin and eventually turning the system into a quantum spin-hall(QSH)insulator. Kane and Mele **[2,3,4]** were first to suggest the existence of QSH effect for pure graphene when moderate to large SOC is taken into account. Apart from the sub-lattice-resolved, giant intrinsic SOCs ( $\Delta^{A}_{soc}, \Delta^{B}_{soc}$) due to the hybridization of the carbon orbitals with the d-orbitals of the transition metal, the other substrate-induced interaction**[5,6,7]** are the extrinsic Rashba spin-orbit coupling (RSOC) ($\lambda_{R}$ $(E)$) that allows for the external tuning of the band gap in Gr-TMD and connects the nearest neighbors with spin-flip, and the orbital gap($\Delta$) related to the transfer of the electronic charge from Gr to TMD. These interactions are absent in isolated pristine, pure graphene monolayer. With these cavalcade of the interactions, the effective, low-energy, dimensionless Hamiltonian for Gr-TMD around the Dirac points $K$ and $K'$ in the basis ($a_{k,\tau_z,\uparrow}$, $b_{k,\tau_z,,\uparrow}$, $a_{k,\tau_z,\downarrow}$, $b_{k,\tau_z,\downarrow}$ ), where $a_{k,\tau_z,s}$ ( $b_{k,\tau_z,s}$ ) is the fermion annihilation operator for the momentum($k$)-valley($\tau_z$) - spin ($s$) state corresponding to the sub-lattice A(B), may be written down as $H_0$ = $[(\tau_z\ ak_x\sigma_x + ak_y\sigma_y) + \Delta\ \sigma_z + \lambda_R\ (E)\ (\tau_z\sigma_x\ s_y - \sigma_y\ s_x) + (\tau_z/2)\{\Delta^{A}_{soc}\sigma_z(s_z+s_0) + \Delta^{B}_{soc}\sigma_z(s_z - s_0)\}]$. The Pauli matrices $\tau_i$, $\sigma_i$, and $s_i$ respectively, are affiliated with the valley degrees of freedom, the pseudo-spin, and the real spin of the Dirac electronic states. We shall replace below $\tau_z$ by its eigenvalue $\xi = \pm1$ for $K$ and $K'$ . Here the nearest neighbor hopping is parameterized by a hybridization $t$, and $\hbar v_F/a = (\sqrt{3}/2)t$. The terms present in the Hamiltonian are made dimensionless dividing by the energy term ($\hbar v_F/a$). Since the Rashba coupling in graphene is constant for mass-less Dirac electrons and does not depend on the momentum components ($\delta k_x$, $\delta k_y$) , the intrinsic RSOC, which is modelled as $\alpha$ ( $\delta k_y\sigma_x - \delta k_x\sigma_y$) (where $\sigma$'s are the Pauli matrices) in conventional semiconducting 2D electron gases, have not been considered here. The graphene layer in GrTMD system is also exchange coupled ($M$) to localized magnetic impurities (MIs), such as substitutional Co atoms, breaking the time-reversal symmetry (TRS). The MIs do not act as scatterer in the present scheme; their effect is included in the band dispersion. The exchange coupling can assist significantly the spin polarization due to the large shift of spin-up and down bands. The gapped, spin-split bands obtained from $H_0$ involve a RSOC-dependent pseudo Zeeman field ( the term $s\sqrt{(z_0(M)/2)}\ \lambda_R$ in Eq.(1) below where $s$ is the spin- chirality index and $\lambda_R$ is the dimensionless RSOC. The Zeeman field here couples with same sign but different manner to the states in the valleys due to $z_0(M)$ being a decreasing function of $M$ for the valley $\mathbf{K}$ and increasing function for the valley $\mathbf{K'}$ [see Figures 1 and 2 inI]. This field encourages the spin precession due to the effective magnetic field in the system over the spin-flip scattering of electrons due to momentum scattering. The field, thus, enables one to obtain an expression of the dielectric function in the finite doping case ignoring the spin-flip scattering events completely. On account of the strong spin-orbit coupling, as already mentioned, the dispersions are gapped at all possible exchange field values with avoided crossings **[8]**.

The surface plasmons (SPs), are defined as quanta of longitudinal in-phase oscillation of all the carriers driven by the (self-consistent) electric field generated by the local variation in charge density at the surface of plasmonic materials. Owing to the heavy energy loss, plasmon inside the materials evanesces severely, but fortunately, it can propagate sufficiently long distance along the surface under suitable condition, such as the increase of doping concentration in the Gr-TMD system (see sub-section 3.2). These collective density oscillations, though can be excited in the conventional metals, suffer large energy losses, such as Ohmic loss and radiative loss. Moreover, SPs in metals have bad tunability in a device**[9]**. On a quick side note, the low losses, the efficient wave localization up to mid-infrared frequencies, the carriers density control by electrical gating and doping**[10]**, and the two dimensional (2D) nature of the collective excitations leading to stronger SP incarceration compared to those in metals make graphene to be a promising plasmonic material. The intra-band( Dirac) Plasmons in a doped graphene sheet are distinctively different($n^{1/4}$ dependence) from that ($n^{1/2}$ dependence) of the conventional 2D electron gas plasmons with respect to the carrier density ($n$) dependence. The former as well as the latter ones exhibit $q^{1/2}$ dependence as is well-known**[11]**.The broad reviews on graphene plasmonics with particular emphasis on the excitations in epitaxial graphene and on the influence of the underlying substrate in the screening processes could be found in refs.**[12,13,14]**. The most general expression of the dynamical polarization at finite temperature, chemical potential, impurity rate, quasi-particle gap, and magnetic field was presented by Pyatkovskiy and Gusynin **[15]** several years ago. As regards Gr-TMD system, starting with $H_0$, the results for the intra-band Plasmon were found to be rather remarkable in I. Since with hole doping the Fermi surface shifts to a lower energy, as a consequence the inter-band transitions with transition energy below twice the Fermi energy become forbidden. It leads to a decrease in higher frequency inter-band (optical) absorption. At the same time, the lower frequency (far infrared and terahertz (THz))free carrier absorption (i.e. intra-band transition) increases dramatically. Therefore, on ignoring the spin-flip mechanism completely for simplicity, the intra-band transitions are needed to be considered only. For these transitions, since the focus is on the long wavelength regime, the transitions between two Dirac nodes located at different momentum could be neglected. These assumptions led to a single collective mode (photon frequency $\hbar\omega \sim (aq)^{2/3}$ where $a$ is lattice constant of grapheme and $q$ is wave vector) in I , corresponding to THz frequencies, for the Gr-TMD system for a finite chemical potential within the RPA by finding the zeros of the dielectric function including only the intra-band transition. In this paper, we have identified the region in the $\frac{\hbar\omega}{\varepsilon_F}$ - $\left(\frac{q}{k_F}\right)$ plane ( here $\boldsymbol{\varepsilon_F}$ is Fermi energy and $k_F$ is Fermi momentum ) where these intra-band plasmons can propagate without damping and are gate-tunable.

The response of a system to electromagnetic waves with wavelength much longer than the characteristic length scales of the systems is given by optical conductivity. The real part of the optical conductivity describes the dissipation of the electromagnetic energy in the medium, while the imaginary part, describes screening of the applied field. We have calculated this response function ($\sigma_0^{intraband}$ and $\sigma_0^{interband}$) for the intra-band and inter-band transitions in this paper. While the real part of the inter-band optical conductivity varies (approximately) linearly with relative to frequency $\omega$ in the limited, low photon energy range as we shall see below in Figure 5, Eq.(10) indicates that the same is not true for the real part of the intra-band optical conductivity; $Re\sigma_0^{intraband} \propto (\hbar\omega)^{-1}$ for a given gate voltage $V_g$ for the frequencies close to the plasmonic resonance. The possible dependence on the gate voltage arises from the dependence on the chemical potential of the fermion number. The changes in the conductivity of Gr-TMD heterostructures from the visible to terahertz region is thus quite remarkable. As regards the Plasmon confinement in Gr-TMD heterostructure, we have shown that possible confinement measure of the surface wave is approximately one-third of the corresponding vacuum wavelength. In the case of graphene, the confinement measure can never be close that of Gr-TMD. It will be even worse if we assume the absence of scattering. We find that there are no (inter-band) plasmons for electron doping as long as the Gr-TMD hetero-structure is surrounded by the positive dielectric constant (ε)materials above and below. However, if $\varepsilon < 0$, i.e. the heterostructure is surrounded by negative dielectric constant (NDC) material which is an essential key for creating meta-materials, we find that there would be acoustic plasmons. For the hole doping case NDC material is not required.

The paper is organized as follows: In section 2, as a recapitulation exercise, we present Gr-TMD band dispersion in I and the dynamical polarization function in random phase approximation (RPA), taking into consideration the general form of low-energy Hamiltonian of graphene monolayer on two dimensional transition metal dichalcogenides Hamiltonian accounting for interactions with the substrate. The exchange field is additionally introduced. In section 3 we obtain the region in the $\frac{\hbar\omega}{\varepsilon_F}$ - $\left(\frac{q}{k_F}\right)$ plane where the intra-band plasmons in I can propagate without damping. We also calculate the intra-band optical conductivity and touch upon the plasmon confinement issue. In section 4, we find the real and the imaginary parts of the inter-band optical conductivity. In section 5, we calculate the transmittance ($T$), at normal incidence. The sub-section 5.1 contains a demonstration that, in the present frame-work, it is not difficult to show that the intra-band acoustic plasmons are a real possibility when the carrier density(Fermi momentum) is large (carrier concentration $n \geq 1.0\times10^{17}\ \mathrm{m}^{-2}$ ) . A potential application of the graphene-transition metal dichalcogenide (TMD)-graphene hybrid nanostructures is the plasmonic meta-surface for ultra-sensitive bio-sensing. An important sensor performance parameter is the reciprocal of full width at half maximum (FWHM). We find that FWHM is narrow at high photon energy and low gate voltage. The paper ends with concluding remarks in section 6.

## 2. Gr-TMD band dispersion and dynamical polarization function

In this section, for the convenience of the reader, we recall some of the important points in I and ref. **[8]**. For example, starting with $H_0$ as in I, we obtain the band dispersion in the form

$$\varepsilon_{\xi,s,\sigma}(ak,M) = [s\sqrt{(z_0(ak,M)/2)}\lambda_R + \sigma\{(ak)^2 + \lambda^2_{\xi,s}(ak,M)\}^{1/2}], \qquad (1)$$

where

$$\lambda_{\xi,s}(ak,M) = \{\beta^2_\xi(M) - z_{0\xi}(ak,M)/2 + s\sqrt{(2c^2_\xi(M)/z_{0\xi}(ak,M))}\}^{1/2},$$

$$\beta_\xi(M) = [(a_1^2 + a_2^2 + a_3^2 + a_4^2)/4 + 4\lambda_R^2]^{1/2}, \quad z_{0\xi}(ak,M) \approx b_\xi(ak,M) + \sqrt{\{d_\xi(ak,M)\}},$$

$$b_\xi(ak,M) = [(a_1^2 + a_2^2 + a_3^2 + a_4^2)/4 + 4\lambda_R^2 + (ak)^2],$$

$$c_\xi(M) = (|\Delta^A_{soc}| - \Delta^B_{soc})\ [2\Delta\{|\Delta^A_{soc}| + \Delta^B_{soc}\} + 4\lambda_R^2 - 4M\Delta\xi],$$

$$d_\xi(ak, M) = [a_1 a_2 a_3 a_4 - ((ak_x)^2 + (ak'_y)^2)(a_1 a_3 + a_2 a_4) + ((ak_x)^2 + (ak'_y)^2)^2 - \lambda_R^2 (1-\xi)^2 (a_1 a_4) - \lambda_R^2 (1+\xi)^2 (a_2 a_3)],$$

$$a_1 = \Delta + M + \xi\,\Delta^A_{soc},\ a_2 = -\Delta + M - \xi\,\Delta^B_{soc},\ a_3 = \Delta - M - \xi\,\Delta^A_{soc},\ a_4 = -\Delta - M + \xi\,\Delta^B_{soc}. \quad (2)$$

The band structure consists of two spin-split valence bands and two spin-split conduction bands. However, since the spin is not conserved because of the spin-flip inducement by the Rashba coupling, the spin is not a good quantum number. The resulting angular momentum eigenstates may be denoted now by the mixed index which, here-in-after, will be referred to simply as 'spin' index $s = \pm 1$. Here $\sigma = +$ $(-)$ indicates the conduction (valence) band. The gapped, spin- valley split bands involve a RSOC-dependent pseudo Zeeman field ( the term $s\sqrt{(z_{0\xi}/2)}\ \lambda_R$ in Eq.(1) )which couples with different signs to states with ↑,↓ spins.

In the random phase approximation (RPA), we write the dynamical polarization function $\chi\,(a\boldsymbol{q}, \omega')$ in the momentum space, as

$$\chi\,(a\boldsymbol{q}, \omega') = \sum_{\xi,\,s,s',\sigma,\sigma'=\pm 1} \chi_{\xi,\,s,s',\sigma,\sigma'}(a\boldsymbol{q}, \omega')$$

$$\chi(a\boldsymbol{q}, \omega') = g_v \sum_{\delta k,\,s,s',\,\sigma,\sigma'=\pm 1} |\langle \Psi_{s,\sigma}(a(\boldsymbol{k}-\boldsymbol{q}), M) \mid \Psi_{s',\sigma'}(a\boldsymbol{k}, M)\rangle|^2$$

$$\times\left[\frac{n_{s,\sigma}(a\boldsymbol{k} - a\boldsymbol{q}, M) - n_{s',\sigma'}(a\boldsymbol{k}, M)}{\{\hbar\omega' + \varepsilon_{\xi=1,s,\sigma}(a(\boldsymbol{k}-\boldsymbol{q}), M) - \varepsilon_{\xi=1,s',\sigma'}(a\boldsymbol{k}, M) + i\eta\}}\right], \quad (3)$$

where $(\hbar\omega/\hbar v_F a^{-1}) = \hbar\omega'$. As was mentioned in I, there are two main problems of applying RPA to real electronic systems, viz. it fails to reduce to the well-known results of classical Boltzmann theory in the long-wavelength limit and it faces the $aq = 2\,ak_F$ non-analyticity problem in the opposite limit. We, however, worked in the long-wavelength limit and thus did not encounter the $aq = 2\,ak_F$ singularity problem. Furthermore, with hole doping the Fermi surface shifts to a lower energy. As a result the inter-band transitions with transition energy below $2E_F$ become forbidden, and it leads to a decrease in higher frequency inter-band absorption. At the same time, the lower frequency free carrier absorption (i.e. intra-band transition) increases dramatically. For this reason we focussed on the long wavelength intra-band plasmons in I. This allowed us to neglect transitions between two Dirac nodes located at different momenta. Within this assumption, contributions from other Dirac node in (3) had been taken into account replacing $\sum_\xi$ by the degeneracy factor $g_v = 2$ and putting the valley index $\xi =$ $+1$ in the summand. The $\varepsilon$ and $\psi$ are single-particle energies and wave funtions, and $n_{\xi,s,\sigma}(a\boldsymbol{k}, M) =$ $[exp(\beta(\varepsilon_{\xi,s,\sigma}(a\boldsymbol{k}, M) - \mu')) + 1]^{-1}$ is occupation function for the band $\sigma = \pm 1$. The spin degenerate band-overlap of wave functions is given by $F_{\sigma,\sigma'}(\delta\boldsymbol{k}, \delta\boldsymbol{q}) = |\langle \Psi_{\xi,\sigma}(a\delta\boldsymbol{k} - a\delta\boldsymbol{q}, M) \mid \Psi_{\xi,\sigma'}(a\delta\boldsymbol{k}, M)\rangle|^2 = (1/2)(1+$ $\sigma\sigma'\cos\theta)$. The angle $\theta$ is that between states at $a(\boldsymbol{k}-\boldsymbol{q})$ and $a\boldsymbol{k}$. For the graphene dispersion $E^{\pm}(\boldsymbol{k}) = \pm t$ $|\phi_{\boldsymbol{k}}| - \mu$ with $\phi_{\boldsymbol{k}} = [1 + 2exp\,(i3ak_x/2)\cos(\sqrt{3}ak_y/2)]$, this overlap of wave functions assumes the form $F_{\sigma,\sigma'}(\boldsymbol{k}, \boldsymbol{q}) = ½[1 + \sigma\sigma' Re(exp(iaq_x)\frac{\phi_{\boldsymbol{k}-\boldsymbol{q}}\,\phi^*_{\boldsymbol{k}}}{|\phi_{\boldsymbol{k}-\boldsymbol{q}}||\phi_{\boldsymbol{k}}|})]$. We assumed near zero temperature scenario for the moment. The occupation function $n_{\xi,s,\sigma}$ for $T \neq 0$ K turns into a simple step function $\theta(\mu' - \varepsilon_{\xi,s,\sigma}$ $(a\delta\boldsymbol{k}, M))$ if T = 0 K. The variable $\mu' = \mu/(\hbar v_F/a)$ is the dimensionless chemical potential of the fermion number. All states below $\mu$ are occupied. If the sum over states k is understood as an integral over all one-particle states, the fermion density in d = 2 dimensions is $n \approx \frac{(k_F)^2}{\pi}$ , where $k_F$ is the Fermi momentum. In terms of the carrier concentration ($n$), the Fermi momentum in pure graphene is therefore $(ak_F) = a\sqrt{(\pi n)}$. The system, however, is with multi-band gapped energy spectrum given by $\varepsilon_{\xi,s,\sigma}(a|\mathbf{k}|,M) \approx [s\sqrt{(z_0(M)/2)}\lambda_R + \sigma\{(a|\mathbf{k}|)^2 + \lambda_s(\xi,M)^2\}^{1/2}]$. The associated Fermi energy at a given concentration($n$)is $\varepsilon_F \approx (1/2)\sum_s \{a^2k_F^2 + \lambda^2_{\xi=1,,s}(M)\}^{1/2}$ . The Fermi momentum $(ak_F)$ is given by $|ak| \le ak_F = \left(\frac{1}{2}\right)[\sqrt{\{(\mu' + \sqrt{(z_0/2)}\lambda_R)^2 - \lambda_{s=1}(\xi=1,M)^2\}} + \sqrt{\{(\mu' - \sqrt{(z_0/2)}\lambda_R)^2 - \lambda_{s=-1}(\xi=1,M)^2\}}]$. The density of states (per unit area) of the system is given by $D(\varepsilon) \approx (2/\pi)\,(|\varepsilon_{\xi,s,\sigma}(a|\mathbf{k}|, M)| / (\hbar v_F)^2)$. For the graphical representation purpose , we had used the well-known **[16]** general relation between $\mu$ and the gate voltage $V_g$ for a graphene-insulator-gate structure, viz.$\mu \approx \varepsilon_a[(m^2+2eV_g/\varepsilon_a)^{1/2} - m]$ where $m$ is the dimension-less ideality factor and $\varepsilon_a$ is the characteristic energy scale. The relation between $\mu$ and the carrier density may be given by $\mu \approx \hbar v_F\sqrt{(\pi|n|)}sgn(n)$ where $sgn(n) = \pm 1$ for electron(hole) doping. The discussion presented below is based on these results.

## 3. Intra-band Plasmon and optical conductivity

We begin by recalling the (Drude model) intra-band optical conductivity expression for the highly doped graphene: $\sigma_0^{intraband}(\boldsymbol{\omega}') \sim (\tau' e^2 k_F/\pi\hbar)\,(1 + i\hbar\omega'\tau')/(1 + (\hbar\omega'\tau')^2)$, where $\tau'^{-1} = \tau^{-1}/\hbar v_F a^{-1}$. One may notice that in the absence of scattering ($\tau'^{-1} << 1$), the expression corresponds to a purely imaginary quantity, signifying no dissipation of electric energy within graphene,though this is unlikely to happen due to the inevitable phonon scattering and finite impurity doping. Furthermore, this expression yields the dispersion relation $\hbar\omega'^{Gr}_{pl} = \hbar\omega'_0(a|\boldsymbol{q}|)^{\frac{1}{2}}$ for the doped graphene in the long wave length limit, where $\hbar\omega'_0 = [(e^2k_F/(2\pi\varepsilon_0\varepsilon\,\hbar v_F\,a^{-1})]^{1/2}$. One can now see that $\omega^{Gr}_{pl} \propto n^{1/4}$ for monolayer graphene which is different from the classical 2D plasmons with $\omega^{2D\,gas}_{pl} \propto n^{1/2}$ though they both have the same dispersion. For the graphical representation purpose, this relation can be expressed as $\hbar\omega^{Gr}_{pl}/\varepsilon_F = [2e^2/(4\pi\varepsilon_0\varepsilon\,\hbar v_F)]^{1/2}\left(\frac{q}{k_F}\right)^{1/2}$ by putting $a^{-1} = k_F$ and the Fermi energy $\varepsilon_F = \hbar v_F k_F$. For $\varepsilon = 2.5$, the dimensionless quantity $[2e^2/(4\pi\varepsilon_0\varepsilon\,\hbar v_F)]^{1/2} \sim 1.35$. For finite value of $q$, the parabolic trend of dispersion relation is seen in Figure 1(a). In the figure we have shown the electron-hole continuum or the single –particle excitation (SPE) region. The region determines the absorption(Landau damping **[17]**) of the external field at a given frequency. We have taken care of the fact that the inter-band transitions with transition energy below $2\,\varepsilon_F$ become forbidden. The plasmons can propagate without damping only in the region (I)not included in the inter-band(II) and the intra-band (III)continuum of SPE. Therefore, the SPs in region I are not damped, however, they will decay into electron-hole pairs above the Fermi wave vector inside the inter-band SPE continuum because of Landau damping **[17]**. Moreover, as $q$ increases, the Drude dispersion relation of graphene plasmon enters the intra-band SPE continuum. These plasmons could be tuned either doping by external potassium atoms **[18]** or rendering graphene gated as could seen from the $\omega^{Gr}_{pl} \propto n^{1/4}$ ( or, $\omega^{Gr}_{pl} \propto \mu^{1/2}$) dependence. The energies of the SPs become higher when the potassium-induced electron density increases. The dispersion relation of low-energy plasmons in a monolayer graphene epitaxially grown on SiC substrate at different carrier density from the High-resolution electron energy loss spectroscopic(EELS) experimental study **[18]**, however, shows that the dispersion relation converges to boundary line of the two kinds of SPE regions but never enters the intra-band SPE continuum. The momentum-dependent EELS spectra is obtained using a parallel beam in transmission electron microscopy (TEM), which reveals the dispersion of plasmons and other spectral excitations of the material.

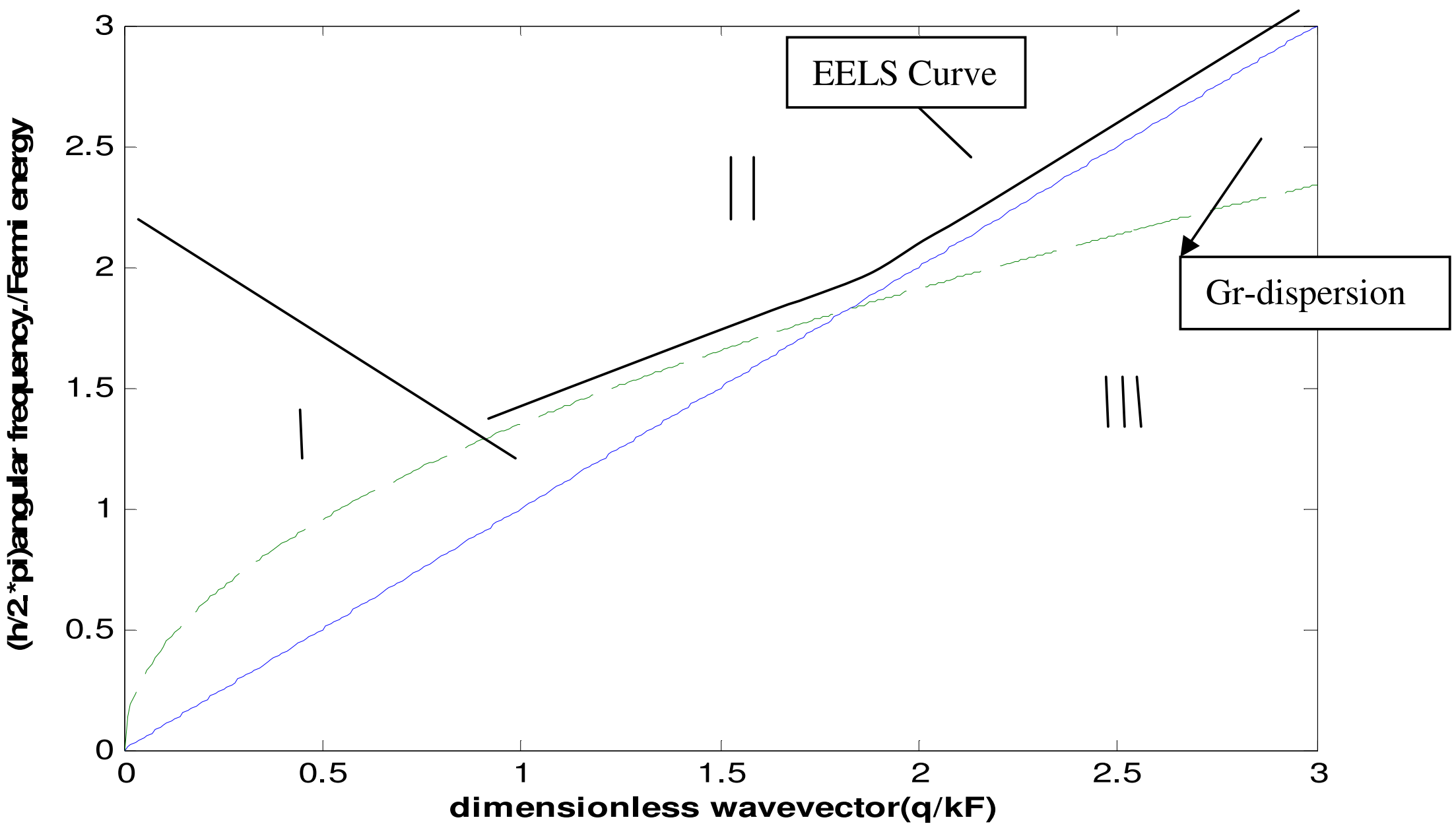

(a)

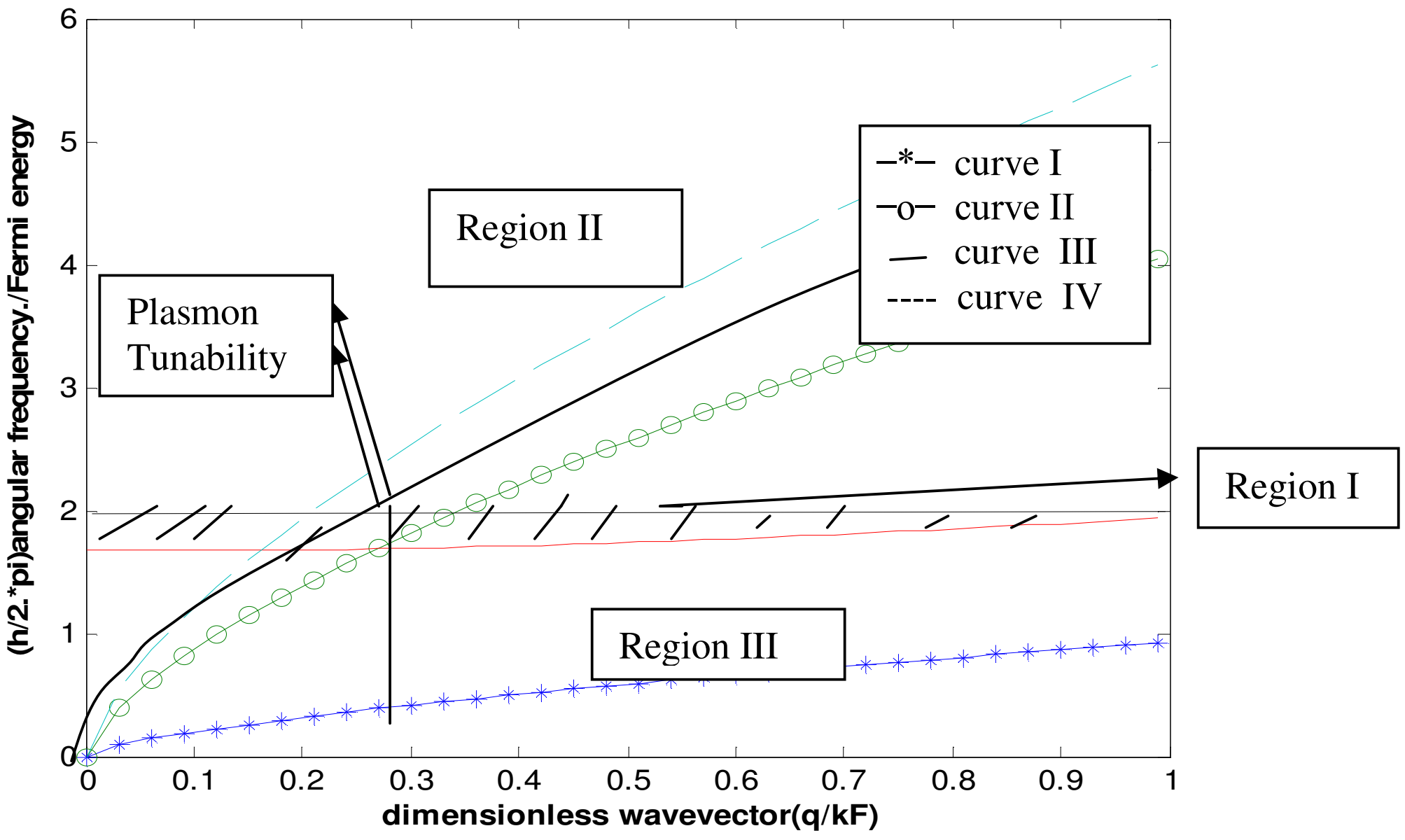

(b)

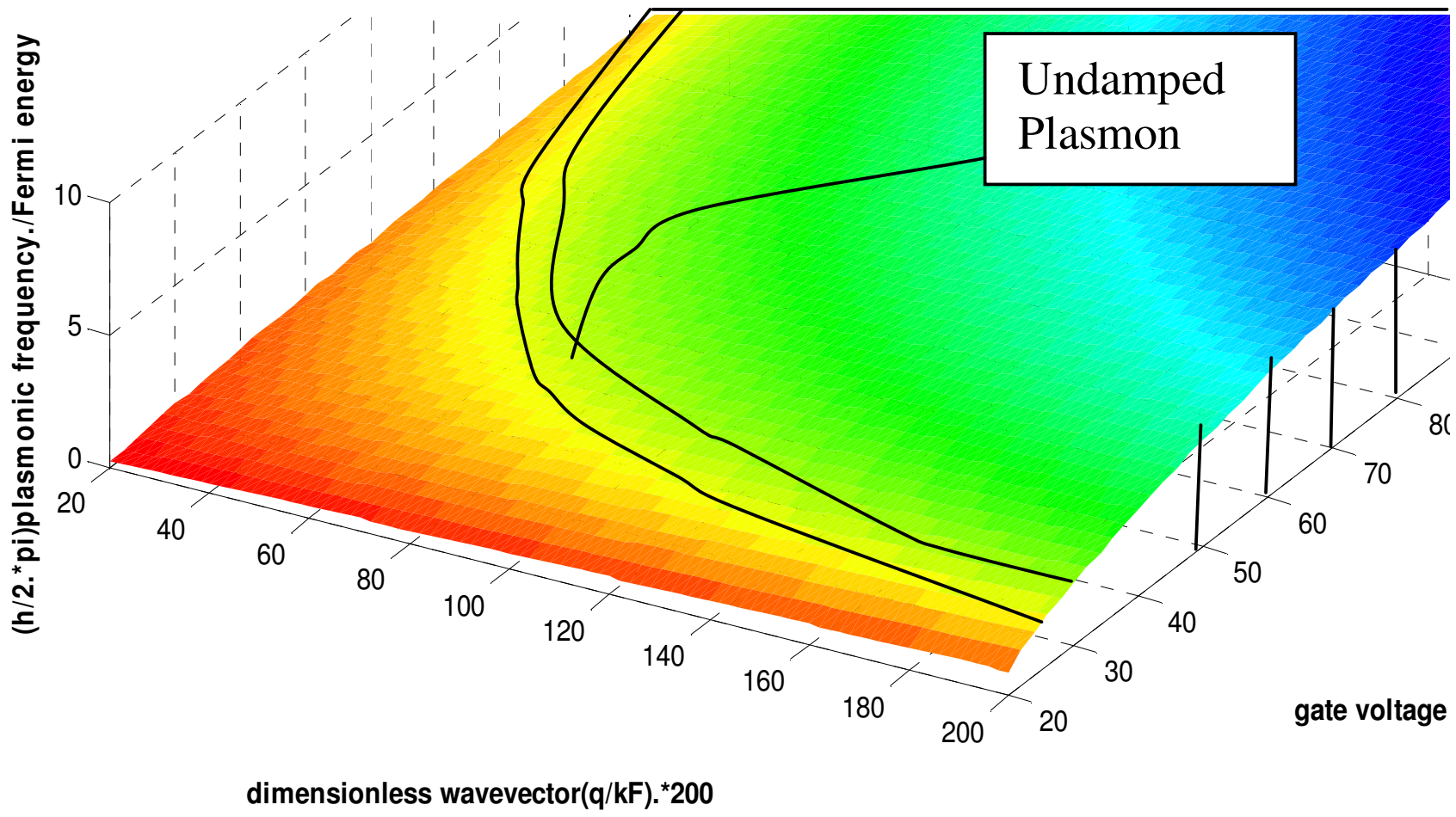

(c)

**Figure 1.(a)** A 2D plot (dashed lines) of $\hbar\omega_{pl}^{Gr}/\varepsilon_F$ as a function of $\left(\frac{q}{k_F}\right)$. The plasmons can propagate without damping only in the region (I)not included in the inter-band(II) and the intra-band (III)continuum of single-particle-excitation (SPE) region, or the electron-hole continuum. The dispersion relation obtained from EELS study, however, shows that the relation converges to boundary line of the two kinds of SPE regions but never enters the intra-band SPE continuum.**(b)** 2D plots of $\frac{\hbar\omega_{pl}^{GrTMD}}{\varepsilon_F}$ as a function of $\left(\frac{q}{k_F}\right)$ for the gate voltage $V_g = 25$ (curve I), 50(curve II), 60( curve III), and 80 volts(curve IV). The hatched region (I) is the region where the plasmons are gate-tunable and can propagate without damping. **(c)** A 3D plot of $\frac{\hbar\omega_{pl}^{GrTMD}}{\varepsilon_F}$ as a function of $\left(\frac{q}{k_F}\right)$ and the gate voltage.

The plasmon dispersion for the Gr-TMD system was obtained in I, within RPA, by finding zeros of the dynamical dielectric function. The dielectric function is expressed as $e_{\xi,s,\sigma}(a\delta\boldsymbol{q},\omega')=1-\frac{V(\delta q)}{\left(\frac{\hbar v_F}{a}\right)}$ $\chi_{\xi,s,\sigma}(a\boldsymbol{q},\omega')$ where $V(q) = (e^2/2\,\varepsilon_0\,\varepsilon_r\,q)$ is the Fourier transform of the Coulomb potential in two dimensions, $V(r) = e^2/4\pi\varepsilon_0\varepsilon_r r$ , and $\varepsilon_0$ the vacuum permittivity and $\varepsilon_r$ is the relative permittivity of the surrounding medium. Upon including the full dispersion of Gr-TMD, ignoring the spin-flip mechanism completely and considering the intra-band transitions only, we have found in I that there is only one collective mode. It corresponds to charge plasmons. For the THz frequency range, as already mentioned, one has to consider the intra-band transitions only. In the long wavelength limit, the corresponding dispersion is given by (see Eq.(21) in I) $\hbar\omega'^{GrTMD}_{pl} = \mathcal{C}^{1/3}\left(a|\boldsymbol{q}|\right)^{\frac{2}{3}} Q^{1/3}(\mu,T,M)$, where $\mathcal{C} = \pi g_v\left(\frac{\frac{e^2}{2a\varepsilon_0\varepsilon_r}}{\left(\frac{\hbar v_F}{a}\right)}\right)$,and

$$Q(\mu,T,M) = \sum_s \int d(a\boldsymbol{k})\,\frac{\beta(a|\boldsymbol{k}|)^3}{4\left|(\lambda_{-s}(\xi=1,M)\right|^3}\cosh^{-2}\frac{\beta}{2}\left(\varepsilon_{\xi=1,s}(a\boldsymbol{k},M)-\mu'\right),\ \beta = (k_B T)^{-1}. \quad (4)$$

Using a representation of the Dirac delta function, viz. $\delta(x)=\mathrm{Lim}_{\grave{e}\to 0}\,(1/2\grave{e}\,cosh^2\frac{x}{\grave{e}})$, Eq.(4) could be reduced to (5) in the zero-temperature limit:

$$Q(\mu,T=0,M) = \left\{\sum_s \frac{\left[\left(\mu'-s\sqrt{\left(\frac{z_0(M)}{2}\right)}\lambda_R\right)^2 - \lambda_{-s}(\xi=1,M)^2\right]^{\frac{3}{2}}}{\lambda_{-s}(\xi,M)^{3}}\right\},\ \mu' = \mu/\hbar v_F a^{-1} \quad (5)$$

The dispersion relation may also be written as

$$\frac{\hbar\omega_{pl}^{GrTMD}}{\varepsilon_F} = [2e^2\, Q(\mu, T=0, M)\,/(4\pi\varepsilon_0\varepsilon\,\hbar v_F)]^{1/3}\left(\frac{q}{k_F}\right)^{\frac{2}{3}} \quad (6)$$

where $\varepsilon_F$ is the Fermi energy given by $\varepsilon_F \approx (1/2)\sum_s \{a^2k_F^2 + \lambda^2{}_{\xi=1,,s}(M)\}^{1/2}$ . The Fermi momentum $(ak_F)$ is given by $|ak| \le ak_F = \left(\frac{1}{2}\right)[\sqrt{\{(\mu' + \sqrt{(z_0/2)}\lambda_R)^2 - \lambda_{s=1}\ (\xi=1,M)^2\}} + \sqrt{\{(\mu' - \sqrt{(z_0/2)}\lambda_R)^2 - \lambda_{s=-1}\ (\xi=1,M)^2\}}]$. For the four Gr-TMD Plasmon dispersion curves in Figure 1(b), we have calculated the quantity $[2e^2\, Q(\mu, T=0, M)\,/(4\pi\varepsilon_0\varepsilon\,\hbar v_F)]^{1/3}$ for the gate voltage $V_g$ = 25, 50, 60, and 80 volts using the general relation **[16]** between $\mu$ and $V_g$, viz.$\mu \approx \varepsilon_a\,[(m^2+2eV_g/\varepsilon_a)^{1/2} - m]$. The fifth curve (red line) is the boundary line of the two kinds of SPE regions. Our dispersion relation for lower voltage, $V_g$ = 25volts as shown in Figure 1(b), never enters the inter-band SPE continuum. It does not converge to the boundary line of the two SPE regions. The plasmons can propagate without damping (and tunable as indicated in Figures 1(b) and 1(c)) only in the region (region I) not included in the inter-band (region II) and the intra-band continuum of SPE (region III).Whereas the tunability of graphene plasmons stems from the $\omega_{pl}^{Gr} \propto n^{1/4}$ ( or, $\omega_{pl}^{Gr} \propto \mu^{1/2}$) dependence albeit the electrostatic/ chemical doping, in the case of the Gr-TMD also it is due to the $\mu$ dependence. The presence of $Q^{1/3}(\mu,T,M)$ term only alters the nature of the dependence. The shrinkage of the region, where the plasmons can propagate without damping, in the case of the Gr-TMD hetero-structure is another point of difference.

### *3.1 Intra-band optical conductivity*

It must be recalled that the dynamical conductivity is the linear response function that relates current density to electric field. The $q \to 0$ ( or, $aq \ll 1$) limit of this is referred to as the optical conductivity because it describes the response of the medium to electromagnetic waves with wavelength much longer than the characteristic length scales of the systems. The real part of the optical conductivity describes the dissipation of the electromagnetic energy in the medium, while the imaginary part, describes screening of the applied field. One obtains the expression for the optical conductivity $\sigma_0^{intraband}(V_g, \boldsymbol{\omega}')$ using the relation $\sigma_{RPA}(q,\omega) = (4i\ \sigma_0\, q^{-2}\hbar\omega)\ \chi^{intraband}(aq,\omega)$, where the real part of the intra-band polarization is

$$\chi_1^{intraband}(a|\boldsymbol{q}|, \omega') = \pi g_v \left(\frac{a|\boldsymbol{q}|}{\hbar\omega'}\right)^3 Q(\mu) \quad (7)$$

(see Eq.(21) in I). The imaginary part, on the other hand, is given by

$$\chi_2^{intraband}(a\boldsymbol{q}, \omega') = -g_v\pi\sum_{\delta k,\, s,s',\, \sigma,\sigma' = \pm 1} [n_{s,\sigma}(a\boldsymbol{k} - a\boldsymbol{q}, M) - n_{s',\sigma'}(a\boldsymbol{k}, M)]$$

$$\times\ \delta\,(\hbar\omega' + \varepsilon_{\xi=1,s,\sigma}(a(\boldsymbol{k}-\boldsymbol{q}), M) - \varepsilon_{\xi=1,s',\sigma'}(a\boldsymbol{k}, M))\ F_{\sigma,\sigma',s,s'}(\boldsymbol{k},\boldsymbol{q}). \quad (8)$$

In writing Eqs.(7) and (8) we have used Eq.(3) and the Sokhotski-Plemelj identity $\{\omega \pm i\eta\}^{-1} = P(\omega^{-1}) \mp i\pi\delta(\omega)$.Now using $\sigma_{RPA}(q,\omega) = (4i\ \sigma_0\ q^{-2}\hbar\omega')\ \chi(aq,\omega)$ and Eqs.(7) and (8) one finds that the imaginary and the real parts of the intra-band dynamical conductivity $\sigma_0^{intraband}$ are

$$Im\sigma_0^{intraband}(q, V_g, \boldsymbol{\omega}') = \mathbf{4}\pi\boldsymbol{\sigma}_\mathbf{0}\ (\hbar\omega')^{-2}\ (a\ q)\ g_v\ Q(\mu, T=0, M),$$

$$Re\sigma_0^{intraband}(q, V_g, \boldsymbol{\omega}') = \mathbf{4}\pi\boldsymbol{\sigma}_\mathbf{0}\ (a\ q)^{-2}\ \hbar\omega' g_v\ K(\mu, \omega', T). \quad (9)$$

The function

$K(\mu, \omega', T)$=$\sum_{,\delta k,s,\sigma}$ $[n_{s,\sigma}(a\boldsymbol{k} - a\boldsymbol{q}, M) - n_{s,\sigma}(a\boldsymbol{k}, M)]$ $\delta$ $(\hbar\omega' + \varepsilon_{\xi=1,s,\sigma}(a(\boldsymbol{k}-\boldsymbol{q}), M) - \varepsilon_{\xi=1,s,\sigma}(a\boldsymbol{k}, M))$

is the reciprocal height of the spectral function for the Gr-TMD (intra-band) plasmon case resembling a Lorentzian (see in section 5 below). In view of Eqs. (15) and (16) in I, it turns out that the function $K(\mu, \omega', T)$ ~ $(\hbar\omega' q_F (V_g, M))$ where $q_F (V_g, M) = \left(\frac{\partial n}{\partial \mu'}\right)$, $n = g_v \sum_{,k,s,\sigma} n_{\xi=1,s,\sigma}(a\boldsymbol{k}, M) F_{\sigma\sigma}(\boldsymbol{k}, \boldsymbol{q})$, and $n_{\xi,s,\sigma}(a\boldsymbol{k},M) = [exp(\beta(\varepsilon_{\xi,s,\sigma}(a\boldsymbol{k},M)-\mu')) + 1]^{-1}$ is Fermi occupation function for the band $\sigma = \pm 1$. We find that zero-temperature limit value of $q_F (V_g, M)$ is given by

$$q_F (V_g, M) \sim \left(\frac{k_F}{\pi}\right)\sum \left\{\frac{1}{\sqrt{[1-\{\frac{\lambda - s(\xi,M)^2}{\left(\mu'(V_g) - s\sqrt{\left(\frac{z_0(M)}{2}\right)}\lambda_R\right)^2}\}]}}\right\}.$$

It has been shown in I that the Thomas-Fermi wave number ($\kappa$) is given by $\kappa = 2\pi\, \alpha\, c\, q_F / (\varepsilon_r v_F)$ where $\alpha = \frac{e^2}{4\pi\varepsilon_0 \hbar c} = \frac{1}{137}$ is the fine-structure constant. Thus, the function $K(\mu, \omega', T) \sim (\hbar\omega' \varepsilon_r v_F \kappa / 2\pi\, \alpha\, c)$. We find from Eq. (9) that, for a given frequency, the real part >> the imaginary part for the Gr-TMD signifying practically no screening of the applied field within the system. Since our plasmon dispersion relation $\hbar\omega_{pl}^{\prime GrTMD} = \mathcal{C}^{1/3} (a|\boldsymbol{q}|)^{\frac{2}{3}} Q^{1/3}(\mu,T,M)$ has been obtained in the long wavelength limit, using this we eventually obtain the real and imaginary parts of the intra-band optical conductivity as

$$Re\sigma_0^{intraband}(V_g, \boldsymbol{\omega}') = \mathbf{4}\pi\sigma_{\mathbf{0}}\, g_v\, (\hbar\omega')^{-1} q_F (V_g, M)\, \mathcal{C}\, Q(\mu, T= 0, M),$$

$$Im\sigma_0^{intraband}(V_g, \boldsymbol{\omega}') = \mathbf{4}\pi\sigma_{\mathbf{0}}\, g_v\, (\hbar\omega')^{-1/2}\, (\mathcal{C}\, Q(\mu, T= 0, M))^{-1/2}. \qquad (10)$$

While the real part of the inter-band optical conductivity varies (approximately) linearly with relative to $\omega$ in the limited, low photon energy range as we shall see below in Figure 5, Eq.(10) indicates that the same is not true for the real part of the intra-band optical conductivity; $Re\sigma_0^{intraband} \propto (\hbar\omega')^{-1}$ for a given gate voltage $V_g$ for the frequencies close to the plasmonic resonance. The possible dependence on the gate voltage arises from the dependence on $\mu'$. In Figure 2(a) we have plotted a 2D graph of the real and imaginary parts of the intra-band optical conductivity as a function of the gate voltage($V_g$)for the dimensionless photon frequencies 400 and 500. The real part increases with $V_g$ and the imaginary part decreases. Both of them decrease with the incident photon frequency. Note that the optical absorbance at normal incidence is related to optical sheet conductivity by $A(\omega, V_g) = (4\pi/c)Re\sigma_0^{intraband}(V_g, \boldsymbol{\omega}')$. A 2D plot of this absorbance as a function of the gate voltage($V_g$)for the dimensionless photon frequencies 400 and 500 is shown in Figure 2(b). It decreases with the frequency at a given gate voltage and increases with $V_g$.The changes in the conductivity of Gr-TMD heterostructures from the visible to terahertz region is thus quite remarkable.

For the scattering rate much smaller than the optical frequency($\omega$) incident, one may put Eq. (10) in the usual Drude form**[19]** with the electron scattering time($\tau$) as

$$\tau\,\omega = (\hbar\omega')^{1/2} / (q_F (V_g, M)\, \mathcal{C}^{3/2}\, Q^{1/2}(\mu, T= 0, M)). \qquad (11)$$

Note that for Gr-TMD heterostructures the scattering time is frequency dependent. The reason for the inclusion of scattering is that, since $c/v_F$ ~ 300, the direct absorption of a photon by an intra-band optical transition does not satisfy momentum conservation. To conserve momentum, extra scattering with phonons or defects is required. It is often convenient to introduce the Drude weight $D = \pi\sigma_0 / \tau$, corresponding to the integrated oscillator strength of free carrier absorption. Here $\sigma_0$ denotes the dc conductivity. For graphene the Drude weight assumes the form: $D = e^2 v_F \sqrt{(\pi n)}$, while for Gr-TMD it assumes a completely different form: $D \sim (e^2 v_F / \hbar k_F)\, [(\pi^2/6)(kT/a^{-1}\hbar v_F)^2 + \mu'^2]$ where $ak_F = (1/2)$ $[\sqrt{\{(\mu' + \sqrt{(z_0/2)}\lambda_R)^2 - \lambda_{s=1}(\xi=1,M)^2\}} + \sqrt{\{(\mu' - \sqrt{(z_0/2)}\lambda_R)^2 - \lambda_{s=-1}(\xi=1,M)^2\}}]$. The possible dependence on the gate voltage arises from the dependence on $\mu'$.

### 3.2 Plasmon confinement

The degree of confinement in a plasmonic waveguide is predominantly determined by the exponential decay of the fields outside the plasmonic waveguide. A crucial factor in this context is referred to as the inverse damping ratio, which is a measure of how many oscillations the plasmon makes before it is damped out completely **[20]**. These quantities could be defined appropriately treating wavevector to be a complex variable : $q = q_1 + iq_2$ with $q_2 \neq 0$. The ratio $R_{FOM} = q_1/q_2 = [\partial / \partial q_1(q_1 \, Im\sigma(q_1,\omega))] / Re\sigma(q_1,\omega)]$ , referred to as the inverse damping ratio. This is the first figure-of-merit (FOM)for the Plasmon propagation. From Eqs. (9) and (10) $R_{FOM} = 4(a\, q_1)^3 (\hbar\omega')^{-3} Q(\mu,T,M) H(\mu,\omega',T)$. The function $H(\mu,\omega',T)$ is given by $H(\mu,\omega',T) \sim (\hbar\omega' q_F)$ . We eventually find $R_{FOM} = 4(a\, q_1)^3 (\hbar\omega')^{-2} q_F\, Q(\mu,T,M) \sim 4(a\, q_1)^{5/3} q_F\, \mathcal{C}^{-2/3} Q^{1/3}(\mu,T,M)$, using the plasmon dispersion $\hbar\omega' \sim (aq)^{2/3}$. The second FOM for the plasmon propagation is the propagation distance of a surface-plasmon $L_{FOM} = q_2^{-1} = 4(a\, q_1)^{2/3} q_F\, \mathcal{C}^{-2/3} Q^{1/3}(\mu,T,M)$. It is desirable to have both FOMs as large as possible for general plasmonic applications. Fortunately, this can be achieved simultaneously. In Figure 3 we have contour plotted $R_{FOM}$ as a function of the absolute value of $(aq)$ and the dimensionless chemical potential for M = 0. The plot clearly emphasizes that the number of Plasmon oscillations before getting damped out completely is likely to be larger than unity for accessible gate voltage values. Likewise, the propagation length of plasmons in Gr-TMD increases with the increase of doping concentration in the system.

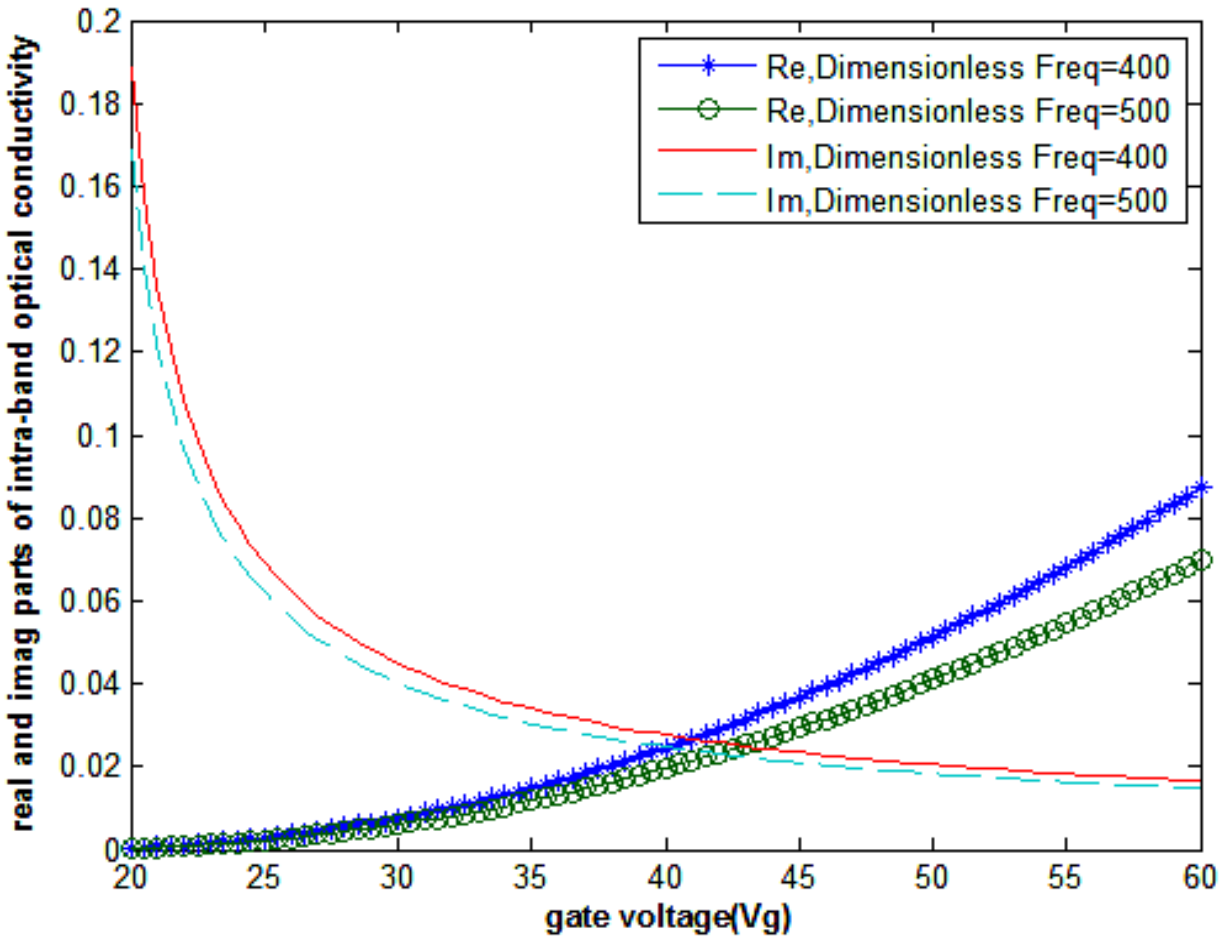

(a)

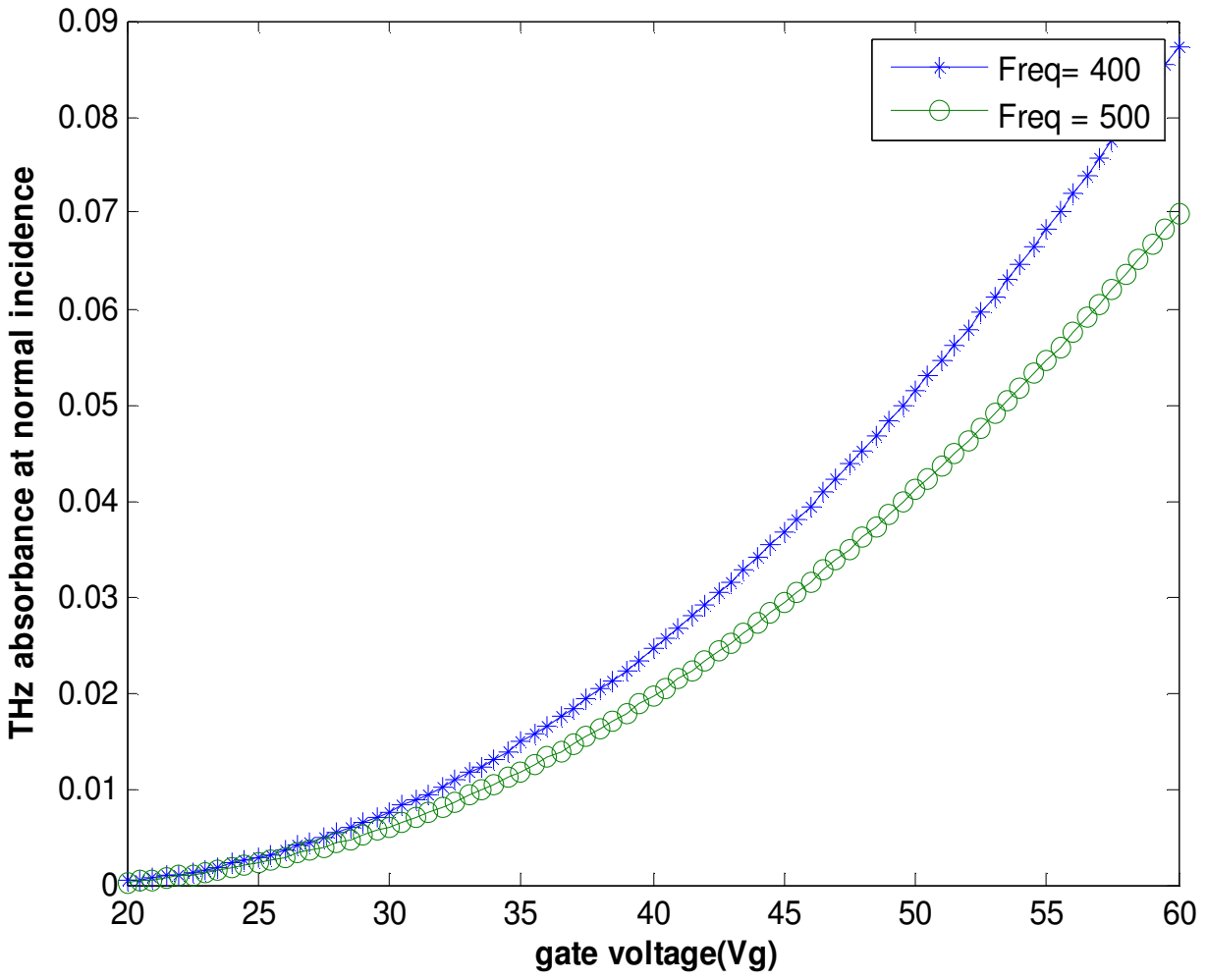

(b)

**Figure 2.(a)** A 2D plot of the real and imaginary parts of the intra-band optical conductivity as a function of the gate voltage($V_g$ )for the dimensionless photon frequencies 400 and 500. The real part increases with $V_g$ and the imaginary part decreases. Both of them decrease with the incident photon frequency. **(b)** A 2D plot of the THz absorbance at normal incidence as a function of the gate voltage($V_g$ )for the dimensionless photon frequencies 400 and 500.

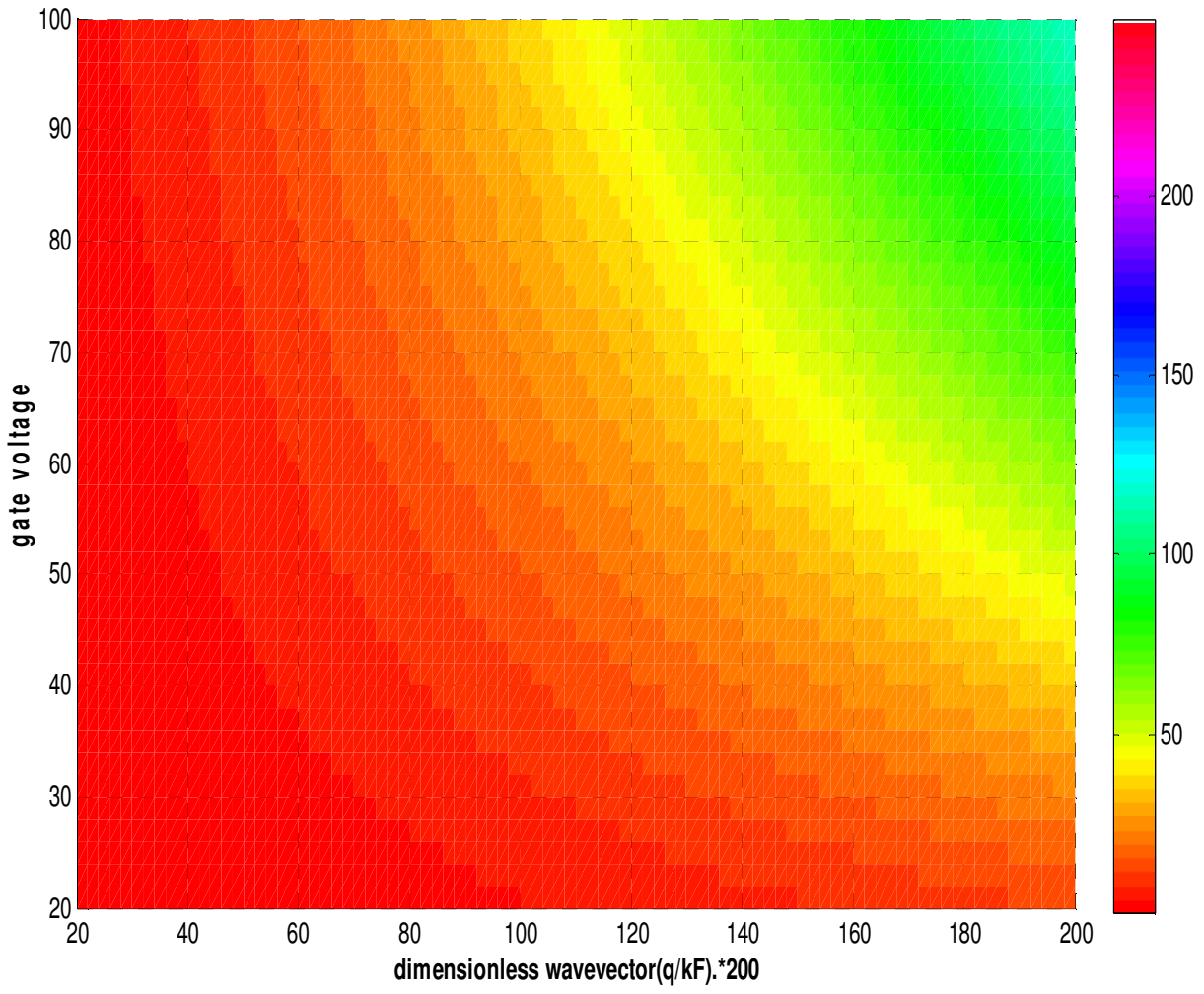

**Figure 3**. A contour plot of the first figure-of-merit (FOM)for the Plasmon propagation as a function of the absolute value of $(aq)$ and the gate voltage in volts for M = 0.

We next need a figure-of-merit which measures the confinement of the surface wave with respect to the corresponding vacuum wavelength. For this we consider the geometry depicted in Figure 4(a), where graphene is surrounded with dielectrics of constants $\epsilon_{r1}$ (TMD) ~ 5 and $\epsilon_{r2}$(Air) =1. We assume that the electric field has the form $E_z = Ae^{iqz-Qx}$ , $E_y = 0$, $E_x = Be^{iqz-Qx}$ , for $x > 0$, $E_z = Ce^{iqz+Q'x}$, $E_y = 0$, $E_x = De^{iqz+Q'x}$ , for $x < 0$. After inserting this ansatz into Maxwells equations and matching the appropriate boundary conditions, we obtain the dispersion relation, in the non-retarded regime ($q \gg \omega/c$), for TM modes as $q \approx Q \approx Q' \approx \epsilon_0\, \epsilon\, 2i\omega/\sigma(\omega, q)$, or $\varepsilon + i(q/2\omega\varepsilon_0)\sigma(q,\omega) = 0$ where the dielectric constant $\varepsilon = (\varepsilon_{r1} + \varepsilon_{r2})/2$, $\varepsilon_0$ is the vacuum permittivity, $\sigma(q,\omega)$ is the wave vector dependent optical conductivity, $q$ is a wave vector, and $\omega$ is the angular frequency of the incident monochromatic optical field. Incidentally, with our expression of $Im\sigma_0^{intraband}(q, V_g, \boldsymbol{\omega}')$ above, upon using this standard equation we obtain $\hbar\omega'_{pl} \propto (a|\boldsymbol{q}|)^{\frac{2}{3}}$.This is a good consistency check; our result is consistent with Maxwells equations. Now an intuitive definition for a figure-of-merit for the degree of Plasmon confinement, on the other hand, is $FOM_{conf} = \lambda_0\, Im(Q)$ where $\lambda_0$ is the free-space wavelength, and $Im(Q)$ is the imaginary part of the complex wave vector $Q$ perpendicular to the interface. This figure-of-merit measures the confinement of the surface wave with respect to the corresponding vacuum wavelength. We obtain in the non-retarded regime $FOM^{Gr\text{-}TMD}_{conf} \sim (\lambda_0/4q_F)(a\, q_1)^{-2/3}\, \mathcal{C}^{2/3} Q^{-1/3}(\mu, T, M)$ . Obviously enough, since the entire body of the calculation above refers to the long wavelength limit $(a\, q_1) << 1$, this confinement measure of the surface wave is larger than one. The presence of the term $Q(\mu, T, M)$ above renders the confinement FOM slightly gate-tunable as shown in Figure 4(b). The figure indicates that maximum possible confinement measure of the surface wave is approximately one-third of the corresponding vacuum wavelength.

In the case of graphene it is not difficult to see, in view of the results $\sigma_0^{intraband}(\boldsymbol{\omega}') \sim (\tau' e^2 k_F/\pi\hbar)\,(1 + i\hbar\omega'\tau')\,/\,(1 + (\hbar\omega'\tau')^2)$ and $q \approx Q \approx Q' \approx \epsilon_0\, \epsilon\, 2i\omega/\sigma(\omega, q)$ ,that the corresponding figure-of-merit for the degree of Plasmon confinement is $FOM^{Gr}_{conf} = \lambda_0\, Im(Q) \sim (2\pi\lambda_0\, \epsilon_0\, \epsilon)\, \hbar\omega/(\tau' e^2 k_F)$. We have to replace here $\hbar\omega$ by $\hbar\omega'^{Gr}_{pl} = \hbar\omega'_0 (a|\boldsymbol{q}|)^{\frac{1}{2}}$ – the Plasmon dispersion relation for the graphene, where $\hbar\omega'_0 = [(e^2 k_F/(2\pi\varepsilon_0 \varepsilon\, \hbar v_F\, a^{-1})]^{1/2}$. This yields in the long wave length limit

$$FOM^{Gr}_{conf} \sim\ = \lambda_0 (a|\boldsymbol{q}|)^{\frac{1}{2}}/(\tau' \hbar\omega'_0).$$

It is clear that the confinement measure of the surface wave in this case can never be close that of Gr-TMD. It will be even worse if we assume the absence of scattering ($\tau'^{-1} << 1$).

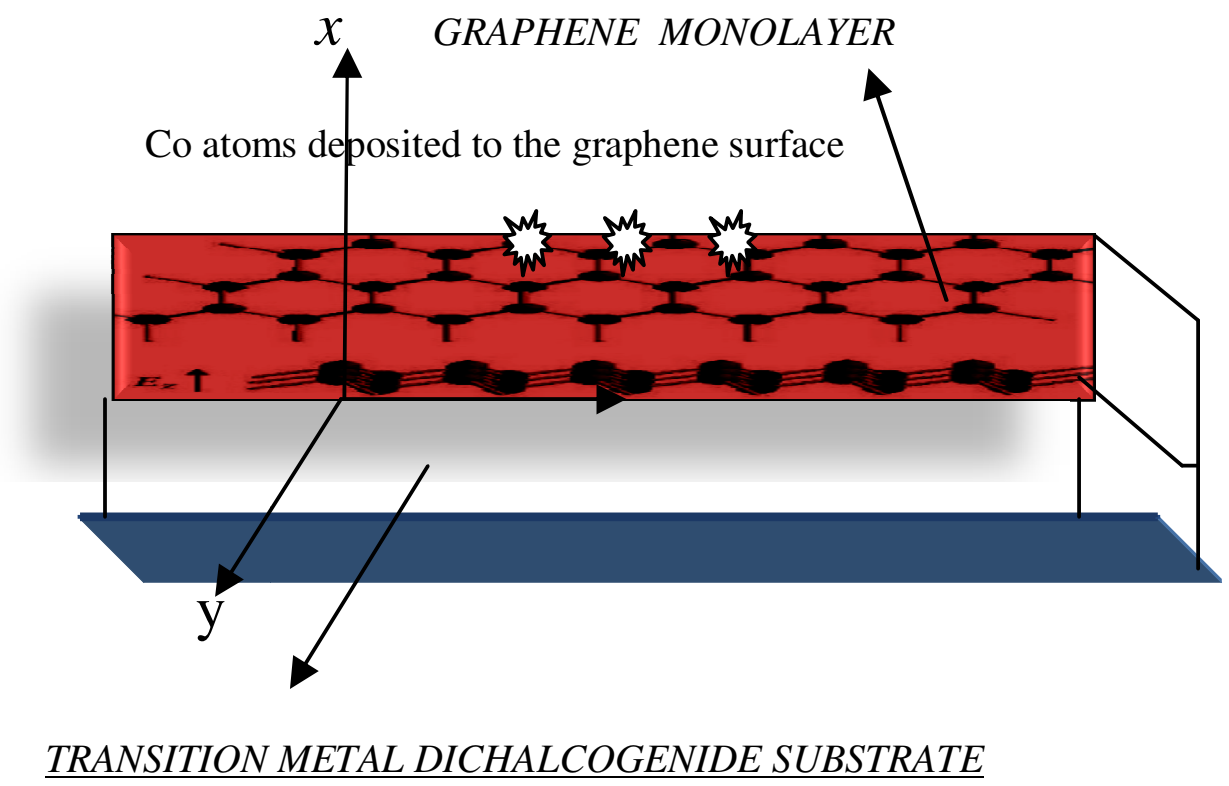

(a)

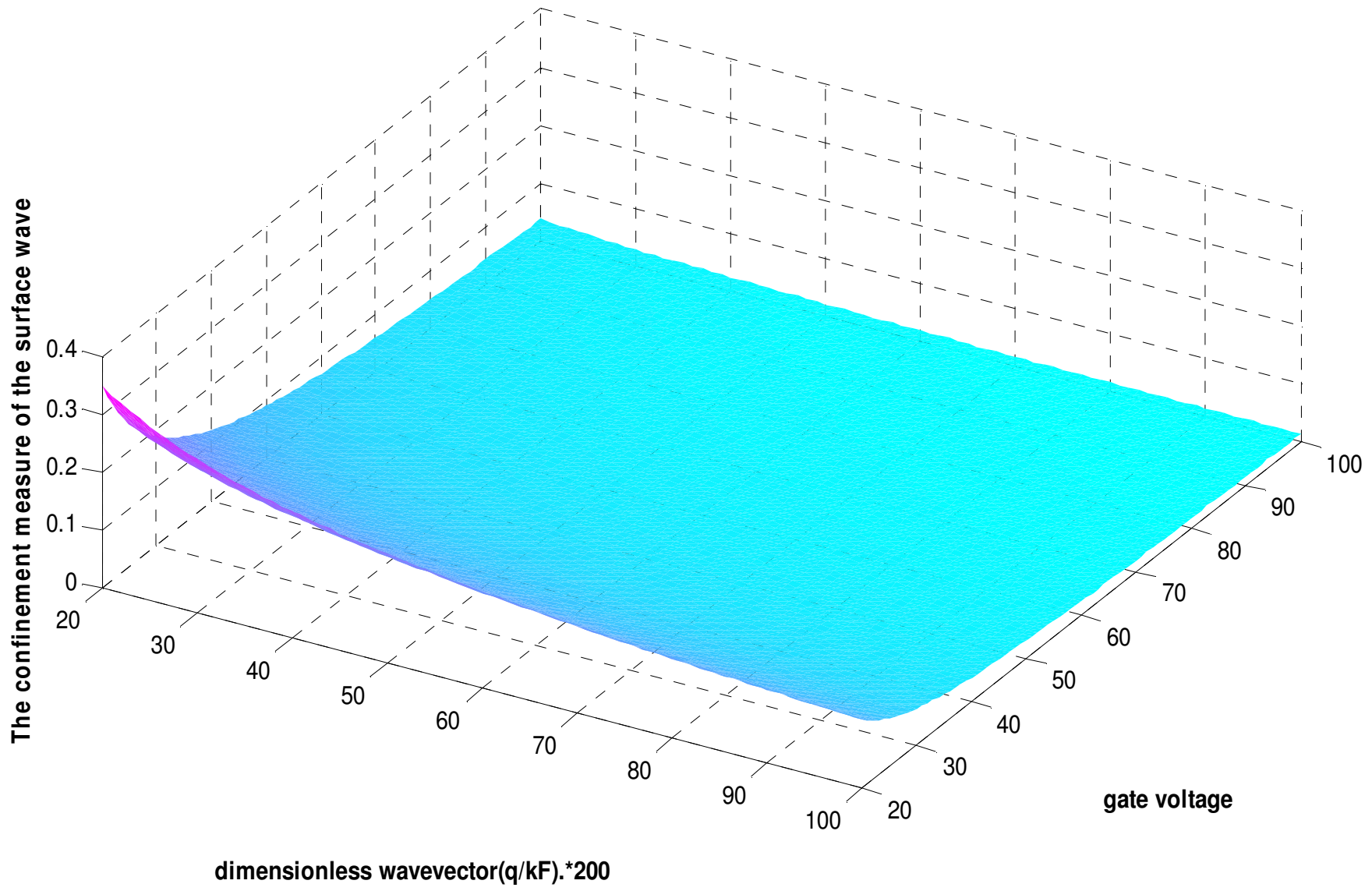

(b)

**Figure 4.** (**a**) A schematic diagram of the graphene on two dimensional TMD. (**b**) A 3D plot of $FOM_{conf}/\,\mathcal{C}^{2/3}\,\lambda_0$ as a function of dimensionless wavevector and the gate voltage.

## 4. Inter-band Optical conductivity

At visible frequencies, inter-band transitions play an important role in most conductors, and a simple Drude model − a free electronic response can no longer represent their optical response. We calculate now the inter-band optical conductivity to understand the role of these transitions. One may note that the perturbation Hamiltonian here is $H_I(t) = -\boldsymbol{\mu_e}.\,\mathcal{E}\,(t)$ where $\boldsymbol{\mu_e}$ is the electric dipole moment and $\mathcal{E}\,(t)$ is a time-dependent electric field. Assuming that the electric field is along x-direction with amplitude $\boldsymbol{E_x}$ and a single frequency ω, one can write $H_I(t) \sim \mu_{e,x}\,\mathrm{Re}[E_x\,exp(-i(\omega+i\eta)t)]$. This ensures that time is negative infinity when the perturbation is turned off. Upon calculating the dipole matrix element $\langle\sigma',\boldsymbol{k}|r_\alpha{}^{\boldsymbol{op}}|\sigma,\boldsymbol{k}\rangle$, of the α-component of the position operator $\boldsymbol{r}_\alpha{}^{\boldsymbol{op}}$ for a transition from the initial state $|\sigma,\,k\rangle$ with energy $\varepsilon_\sigma(\boldsymbol{k})$ into the final state $|\sigma',\,k\rangle$ with energy $\varepsilon_{\sigma'}(\boldsymbol{k})$ as a first step, one obtains the perturbation Hamiltonian $H_I(t)$ in the interaction picture. Here, the indices $\sigma$ and $\sigma'$ denote the spin and all band quantum numbers for the occupied and empty states respectively, $k$ is the continuous quantum number related to the translational symmetry and restricted to the first Brillouin zone. This must be followed by the calculation of the time evolution of the off-diagonal elements of the density matrix $\rho(t)$ given by $i\hbar d\rho(t)\,/\,dt = [H_I(t),\,\rho(t)]$. The third step is to consider the electric current $j_{xx}(t) = -\,(e/A)\mathrm{Tr}(\rho(t)v_x)$, $j_{xx}(\omega) = \sigma\,(\omega)\,\mathcal{E}(\omega)$, where $A$ is sectional area and $v_x$ is velocity operator, and calculate the latter and then the former. One eventually arrives at the relation for the inter-band conductivity in terms of the frequency-independent universal sheet conductivity of the mass-less Dirac fermions is $\sigma_0=(e^2/4\hbar)$,viz.$\sigma_0{}^{interband}(\omega,V_g) = (4i\sigma_0/m_e{}^2\omega\,\hbar\,A)\sum_{\sigma,\sigma',\xi,s,s'}\int(d(ak_x)\,d(ak_y)/(2\pi)^2\,F^{\alpha\beta}{}_{\sigma,\sigma'}(ak)\times[\frac{\boldsymbol{n_{s,\xi,\sigma}(ak},\mu')-\boldsymbol{n_{s',\xi,\sigma'}(ak},\mu')}{\{\hbar\boldsymbol{\omega}'+\varepsilon_{\xi,\boldsymbol{s},\sigma}(\boldsymbol{ak})-\varepsilon_{\xi,\boldsymbol{s}',\sigma'}(\boldsymbol{ak})+i\eta\}}]$, where $m_e$ denotes the electron mass, $\omega$ is the angular frequency of the electromagnetic radiation causing the transition, $(\hbar\,\omega\,/\,\hbar\,v_Fa^{-1}) = \hbar\,\omega'$, $F^{\alpha\beta}{}_{\sigma,\sigma'}(ak) = \prod^\alpha{}_{\sigma,\sigma'}(ak)\prod^\beta{}_{\sigma,\sigma'}(ak)$, $\mu' = (\mu\,/\,\hbar\,v_Fa^{-1})$ is the dimensionless chemical potential of the fermion number, and $\prod^\alpha{}_{\sigma,\sigma'}(\boldsymbol{k}) = \langle\sigma',\boldsymbol{k}|\boldsymbol{p}_\alpha{}^{\boldsymbol{op}}|\sigma,\boldsymbol{k}\rangle$ is the transition matrix element of the α-

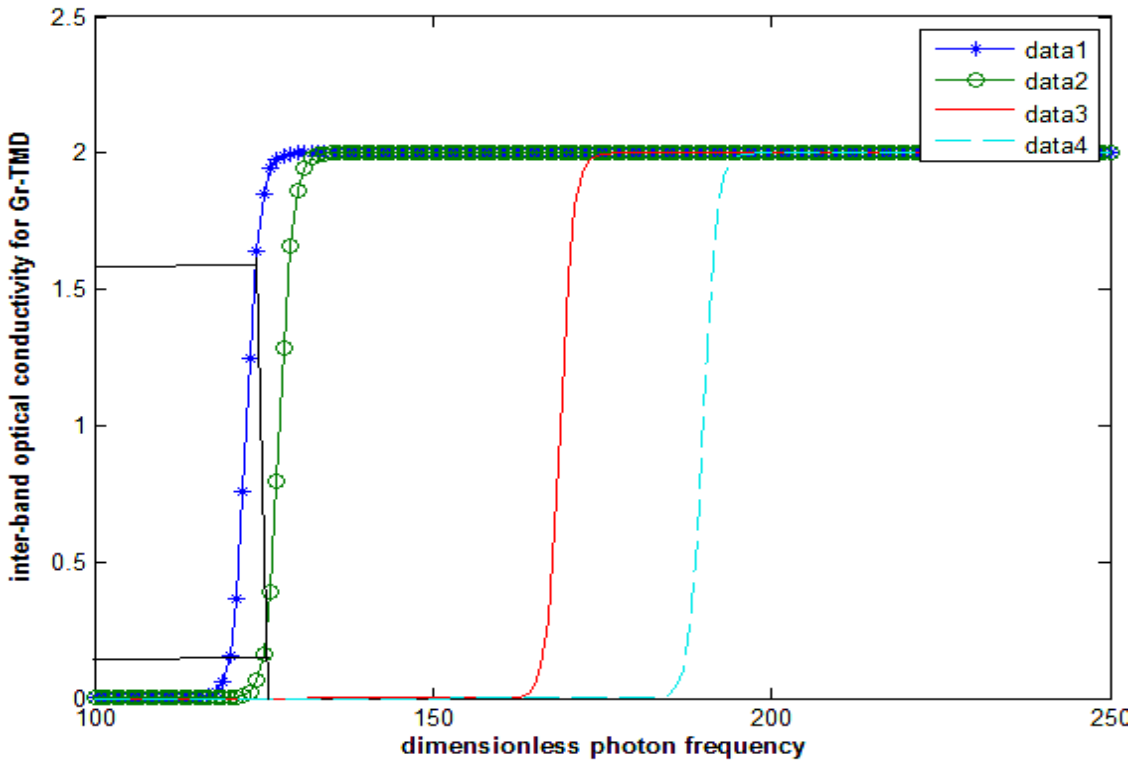

(a)

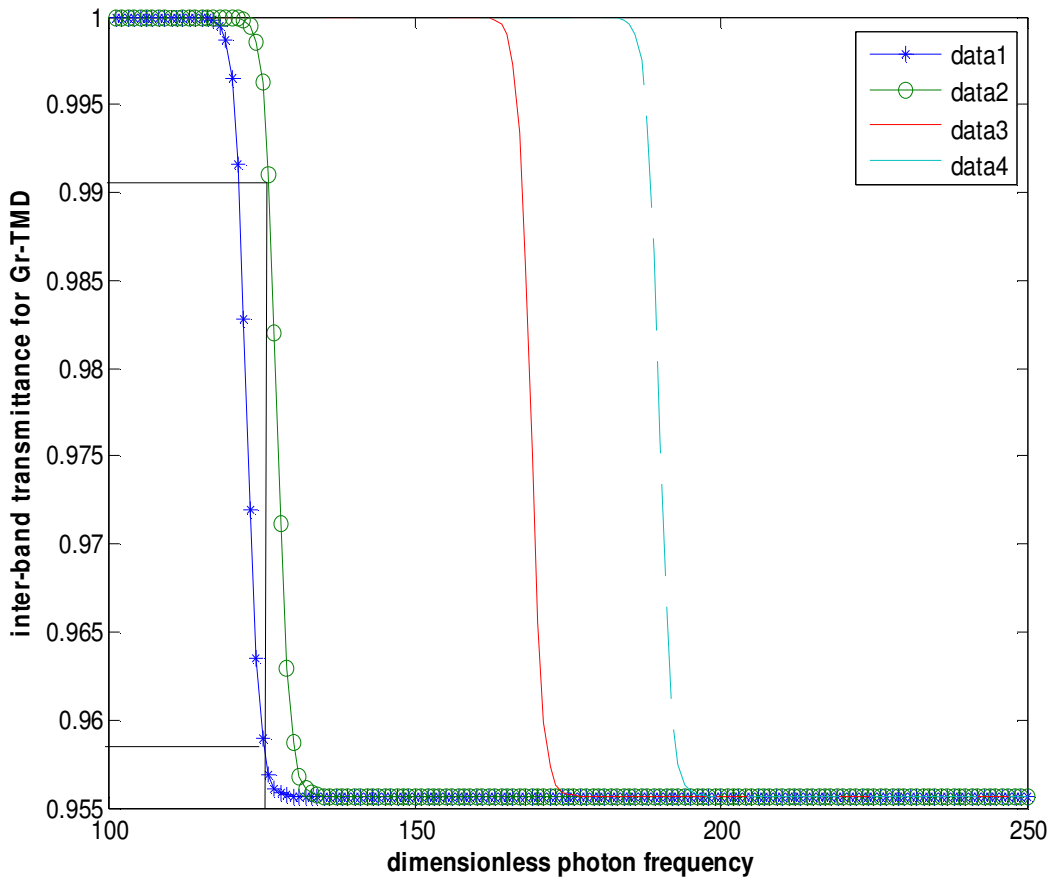

(b)

**Figure 5**. **(a)** A 2D plot of the real part of the inter-band optical conductivity of Gr-TMD, in terms of $\sigma_0=(e^2/4\hbar)$, as a function of the dimensionless photon energy($\hbar\omega/kT$). Here the Boltzmann constant multiplied by the temperature( kT ) = 30 meV . The gate voltage in meV for data 1 is 20, for data 2 it is 21, for data 3 it is 30, and for data 4 it is 35. There is a guide to eye for the curves 1 and 2 to show that the inter-band optical conductivity is tunable. The optical conductivity increases with decrease in the gate voltage. **(b)** A 2D plot of the transmittance of Gr-TMD as a function of the dimensionless photon energy($\hbar\omega/kT$). The transmittance decreases with the decrease in the gate voltage.

component of the momentum operator $\boldsymbol{p}_\alpha{}^{\boldsymbol{op}}$ for a transition from the initial state $|\sigma, k\rangle$ with energy $\varepsilon_{\xi\boldsymbol{s},\sigma}(\boldsymbol{ak})$ into the final state $|\sigma', k\rangle$ with energy $\varepsilon_{\xi\boldsymbol{s}',\sigma'}(\boldsymbol{ak})$. Ignoring the many-body effects, one may write the dimensionless single particle energies ( $\varepsilon_{\xi\boldsymbol{s},\sigma}(\boldsymbol{ak})$, $\varepsilon_{\xi\boldsymbol{s}',\sigma'}(\boldsymbol{ak})$) less than the Fermi mometum $(ak_F)$. The set of indices ($\boldsymbol{k}$, $\xi$, $s$) are for momentum($\boldsymbol{k}$)-valley($\xi = \pm1$ for $\boldsymbol{K}$ and $\boldsymbol{K}'$ ) - spin ($s = \pm1$ ) state corresponding to the graphene sub-lattice A(B). Here, the occupation function for the band $\sigma$ is $n_{\xi s,\sigma}(ak) = [exp(\beta(\varepsilon_{\xi\boldsymbol{s},\sigma}(ak)-\mu')) + 1]^{-1}$ , and $\beta = (\hbar\, v_F a^{-1}/\, k_B T)$. Since graphene is isotropic in $x$ and $y$ directions and the dispersion band is linear, we can replace the average velocity $v_x^2$ with $v_F^2/2$ where $v_F$ is the electronic group velocity of carriers on the Dirac band in graphene. Since for Gr-TMD the dispersion band is different and it is given by

$$\varepsilon_{\xi,s,\sigma}(a|\boldsymbol{k}|, M) = [s\sqrt{(z_0/2)}\,\lambda_R + \sigma\{(a|\boldsymbol{k}|)^2 + \lambda_{-s}(\xi,M)^2\}^{1/2}],$$

this will not be possible. One finds that $v_x = \frac{\partial \varepsilon_{\xi,s,\sigma}}{\partial(ak_x)} = \frac{\sigma(ak_x)}{\{(a|\delta\boldsymbol{k}|)^2 + \lambda_{-s}(\xi,M)^2\}^{1/2}}$. Upon using the Sokhotski-Plemelj identity $\{\omega - i\eta\}^{-1} = \mathrm{P}(\omega^{-1}) + i\pi\delta(\omega)$, after some algebra, one obtains an analytical form for the real part in the high frequency limit ($\hbar\omega > 2\mu$) as

$$\mathrm{Re}\,\sigma_0^{interband}(\omega, V_g) \approx \sigma_0 \sum_{s,\xi} A_{s,\xi}(\hbar\omega, M)\left[\tanh\left(\frac{\hbar\omega+2\mu}{4kT}\right) + \tanh\left(\frac{\hbar\omega-2\mu}{4kT}\right)\right] \quad (12)$$

where $A_{s,\xi}\left(\frac{\hbar\omega}{2kT}, M\right) = \frac{\left[\left(\frac{\hbar\omega}{2kT} - s\sqrt{\left(\frac{z_0(M)}{2}\right)}\lambda_R\right)^2 + \lambda_{-s}(\xi,M)^2\right]}{2\left(\frac{\hbar\omega}{2kT} - s\sqrt{\left(\frac{z_0(M)}{2}\right)}\lambda_R\right)^2}$, in the Dirac cone approximation. We have used the result $[(1/2) - (exp(z) + 1)^{-1}] = (1/2)\, tanh\,(z/2)$ above. The approximation assumes that the conductivity is only contributed by the carriers close to the Dirac cone. The Dirac cone approximation is only valid if the Fermi energy and the photon energy is within the visible range. Beyond that, one can no longer adopt the approximation. One finds $\mathrm{Re}\sigma_0^{interband}(\omega, V_g)$ to be an increasing function of $\omega$, a decreasing function of the chemical potential or the gate voltage $V_g$ (see Figure 5 ). It is, in fact, (approximately) linearly varying with relative to $\omega$ in the limited, low photon energy range. The possible dependence on the gate voltage $V_g$ arises from the dependence on $\mu'$ and the general relation **[16]** between $\mu$ and $V_g$ for a graphene-insulator-gate structure, viz.$\mu \approx \varepsilon_a[(m^2 + 2eV_g/\varepsilon_a)^{1/2} - m]$ where $m$ is the dimension-less ideality factor and $\varepsilon_a$ is the characteristic energy scale. The imaginary part of the optical conductivity $I_0^{interband}(\omega, V_g)$ is given by the first Kramers-Kronig (KK) relation**[17]**, $Im\sigma_0^{interband}(\omega, V_g) = (-2/\pi)\, \boldsymbol{P}\int_0^\infty \omega\, Re\sigma_0^{\mathrm{interband}}(\varepsilon, V_g)\, d\varepsilon/[\varepsilon^2 - \omega^2]$ where $P$ denotes the Cauchy principal value. The second relation gives the real part when the imaginary part is given: $\mathrm{Re}\sigma_0^{interband}(\omega, V_g) = (2/\pi)\, P\int_0^\infty \varepsilon\, Im\sigma_0^{\boldsymbol{interband}}(\varepsilon, \boldsymbol{V_g})\, \boldsymbol{d}\varepsilon/[\varepsilon^2 - \omega^2]$. It is tedious, though straightforward, to show that in the low-temperature regime

$$Im\,\sigma_0^{interband}(\omega, V_g) = (2\sigma_0/\pi)\, \boldsymbol{P}\int_0^\infty \omega\, d\varepsilon \frac{\sum_{s,\xi} A_{s,\xi}(\varepsilon, M)\left[\tanh\left(\frac{\varepsilon+2\mu}{4kT}\right) + \tanh\left(\frac{\varepsilon-2\mu}{4kT}\right)\right]}{(\omega^2 - \varepsilon^2)}$$

$$\sim (2\sigma_0/\pi)\sum_{s,\xi}\left[\frac{\mu}{2(kT)^2} - (2/\hbar\omega)\left\{\frac{(\hbar\omega)^2}{(4kT)^2} + \frac{\lambda_{-s}(\xi,\boldsymbol{M})^2}{4}\right\}\ln\left\{\frac{\hbar\omega+4\mu}{\hbar\omega-4\mu}\right\}\right]. \quad (13)$$

We have noted above that the plasmon dispersion could be obtained in the non-retarded regime ($q \gg \omega/c$) by solving the equation $\varepsilon + i\,\hbar\,(q/2\,\hbar\omega\varepsilon_0)\sigma(q,\omega) = 0$.With our expression of $Im\sigma_0^{interband}$ above, upon using this equation, in the high frequency limit ($\hbar\omega > 4\mu$) and $\mu > 2(kT)$ we obtain

$$\varepsilon\varepsilon_0 + \sigma_0\,(\hbar\, aq/\pi\hbar\omega)\sum_{s,\xi}\left[(2/\hbar\omega)\left\{\frac{(\hbar\omega)^2}{(4kT)^2} + \frac{\lambda_{-s}(\xi,\boldsymbol{M})^2}{4}\right\}\ln\left\{\frac{\hbar\omega+4\mu}{\hbar\omega-4\mu}\right\} - \frac{\mu}{2(kT)^2}\right] = 0. \quad (14)$$

After a little algebra, this equation simplifies to

$$(\hbar\omega')^3 + 4\,C'\mu' aq\,\eta^2(\hbar\omega')^2 + 4\,C'\mu' aq\sum_{s,\xi}\lambda_{-s}(\xi,\boldsymbol{M})^2 = 0, \quad (15)$$

where $C' = C / \left(\frac{\hbar v_F}{a}\right)$ , $C = \pi g_v\left(\frac{\frac{e^2}{2a\varepsilon_0\varepsilon}}{\left(\frac{\hbar v_F}{a}\right)}\right)$ and $\eta = \frac{\left(\frac{\hbar v_F}{a}\right)}{\sqrt{2}\,kT}$. Obviously enough, there are no (inter-band) plasmons for electron doping as long as $\varepsilon > 0$. However, if $\varepsilon < 0$, i.e. graphene on TMD is surrounded by negative dielectric constant (NDC) material which is an essential key for creating meta-materials, we find that there would be acoustic plasmons with dispersion relation $(\hbar\omega') \approx 4\,|C'|\,\mu'\,\eta^2(aq)$. For the hole doping case NDC material is not required; we obtain the acoustic plasmons with dispersion $(\hbar\omega') \approx 4\,C'|\,\mu'\,|\eta^2(aq)$. In the present frame-work it is not difficult to demonstrate (see sub-section 5.1)that the intra-band acoustic plasmons also are a real possibility when the carrier density (Fermi momentum) is large (carrier concentration $n \geq 1.0\times10^{17}\ \mathrm{m}^{-2}$ ).

## 5. Discussion

One sees that if the transmission spectra is given, in principle, it is possible to obtain the real and the imaginary parts of the optical conductivity. As regards the optical absorbance ($A$) and the transmittance ($T$),the latter at normal incidence is given by the equation $T(\omega,V_g) = (1+ Re\sigma_0^{\ interband}(\omega,V_g) / (2c\varepsilon_0))^{-2}$ where $\mathrm{Re}\ \sigma_0^{\ interband}(\omega,V_g)$ is the real part of the inter-band optical conductivity. Upon replacing $Re\sigma_0^{\ interband}$ by $\sigma_0=(e^2/4\ \hbar)$ is the frequency-independent universal sheet conductivity of graphene above, one obtains $T(\omega,V_g) \approx 1-\pi\alpha$ where $\alpha = e^2 / (4\pi\varepsilon_0\,\hbar\, c)$ is the fine structure constant and $c$ is the speed of light. The absorbance of graphene, thus, corresponds to the well-known**[21]** value $\pi\alpha= 2.3\%$. The system under consideration, however, is the gapped graphene (Gr-TMD). The transmittance ($T$) here, at the normal incidence, is given by the relation

$$T(\omega,V_g) = \{1+ (\pi\alpha/2)\sum_{s,\xi} A_{s,\xi}\,(\hbar\omega, M)\,[tanh\,(\frac{\hbar\omega+2\mu}{4kT}) + tanh\,(\frac{\hbar\omega-2\mu}{4kT})]\}^{-2}. \qquad (16)$$

For the transmission of the optical frequencies through the system, only the real part of the inter-band conductivity is relevant. The reason for this is that the thickness of the vdWH in question is several orders of magnitude smaller than an optical wavelength. In view of the variation shown in Figure 5(a), we find that the optical conductivity increases with decrease in the gate voltage. A 2D plot of the transmittance of Gr-TMD as a function of the dimensionless photon energy($\hbar\omega/kT$) in Figure 5(b), however, shows that the transmittance decreases with the decrease in the gate voltage. The optical absorbance is an increasing function of $\omega$ and increases relative to the decrease in $V_g$.

### *5.1 Intra-band acoustic plasmon dispersion.*

To demonstrate the existence of intra-band acoustic plasmons one may consider the dynamical polarization $\chi_{\xi,m}^{(1)}(a\boldsymbol{\delta q},\boldsymbol{\omega}')$ expressed in terms of the first order contribution in the electron–electron interaction in Eq.(14a) in I. The self-consistent RPA result of the polarization to all orders in the electron–electron interaction, in the long wave length limit $(\boldsymbol{aq}) \ll 1$,is given by **[22]** $\chi_{\xi,m}(a\boldsymbol{q},\boldsymbol{\omega}')=\chi_{\xi,m}^{(1)}(a\boldsymbol{q},\boldsymbol{\omega}')/[1-\frac{V(q)}{\left(\frac{\hbar v_F}{a}\right)}\chi_{\xi,m}^{(1)}(a\boldsymbol{q},\boldsymbol{\omega}')] \approx \chi_{\xi,m}^{(1)}(a\boldsymbol{q},\boldsymbol{\omega}')/\,\boldsymbol{e}\,(\boldsymbol{\delta q},\boldsymbol{0}) = \chi_{\xi,m}^{(1)}(a\boldsymbol{q},\boldsymbol{\omega}')/[1+(\kappa/\boldsymbol{q})]$ which leads to the plasmon dispersion obtainable, within the RPA, by finding the zeros of the self-consistent dielectric function $e_{\xi,m}(a\boldsymbol{q},\boldsymbol{\omega}')=1-\frac{V(q)}{\left(\frac{\hbar v_F}{a}\right)}\chi_{\xi,m}(a\boldsymbol{q},\boldsymbol{\omega}')=1-\frac{V(q)}{\left(\frac{\hbar v_F}{a}\right)}[\chi_{\xi,m}^{(1)}(a\boldsymbol{q},\boldsymbol{\omega}')/\{1+(\kappa/\boldsymbol{q})\}]$, where $V(q) = (e^2/2\,\varepsilon_0\,\varepsilon_r\,q)$ is the Fourier transform of the Coulomb potential in two dimensions, $\boldsymbol{v_F}$is the Fermi velocity, $a$ is the lattice constant of graphene, and $\kappa$is the Thomas-Fermi (TF)wave vector. The TF wave vector is given by

$$\kappa \sim e^2 q_F / (2\hbar v_F \varepsilon_0 \varepsilon_r) = 2\pi\,\alpha\, c\, q_F/\varepsilon_r v_F = (2\pi\,\alpha\, c\,/\varepsilon_r v_F)\left(\frac{k_F}{\pi}\right)\sum\left\{\frac{1}{\sqrt{[1-\{\frac{\lambda_{-s}(\xi,M)^2}{\left(\mu' - s\sqrt{\left(\frac{z_0(M)}{2}\right)}\lambda_R\right)^2}\}]}}\right\}. \quad (17)$$

Upon substituting the expression for $\chi_{\xi,m}^{(1)}(aq,\omega')$ from Eq.(20a) in I to the equation above, one obtains the equation $e_{\xi,m}(aq,\omega') = 0$ in the form

$$0 = 1 - \frac{V(q)}{\left(\frac{\hbar v_F}{a}\right) e\,(q,\omega'=0)}\,\pi g_v \left(\frac{a|q|}{\hbar\omega'}\right)^3 \sum_m \frac{\left[\left(\mu' - m\sqrt{\left(\frac{z_0(M)}{2}\right)}\lambda_R\right)^2 - \lambda_{-m}(\xi,M)^2\right]^{\frac{3}{2}}}{\lambda_{-m}(\xi,M)^{-3}}. \quad (18)$$

All the symbols in (17) and (18) are clearly defined in I. In particular, the effective static dielectric constant of the Gr-TMD system is $e\,(q,\omega'=0) = \left\{1 + \left(\frac{\kappa}{q}\right)\right\}$. It is clear from (1) that, in the long wave length limit, for large carrier concentration ($n \geq 1.0\times10^{17}$ m$^{-2}$ ) $\{1 + (\kappa/q)\} \approx (\kappa/q)$ and therefore $\hbar\omega' \sim C(\mu') \times a|q|$ ( acoustic Plasmon dispersion).The tunability of this type of plasmons is also obvious from (1) as $C$ is a function of $\mu'$. The group velocity of these plasmons is about two orders of magnitude lower than speed of light; thus, its direct excitation by light is not possible. The linear behaviour of the dispersion implies that group and phase velocities are the same. So, signals can be transmitted undistorted along the surface. This finding has significant importance in extremely low loss communications. For example, the relation $e\left(q = q_F, \omega' = 0\right) = \left\{1 + \left(\frac{\kappa}{q_F}\right)\right\}$ with $\kappa$ given by (17) indicates that the refractive index of $\mathrm{RI}_{\text{Gr-TMD}} \sim \left\{1 + \left(\frac{\kappa}{q_F}\right)\right\}^{1/2}$ is also tunable. From (17) we find that

$$\mathrm{RI}_{\text{Gr-TMD}} \sim \left[1 + (2\,\alpha\, c\,/\varepsilon_r v_F)\sum\left\{\frac{1}{\sqrt{[1-\{\frac{\lambda_{-s}(\xi,M)^2}{\left(\mu' - s\sqrt{\left(\frac{z_0(M)}{2}\right)}\lambda_R\right)^2}\}]}}\right\}\right]^{1/2}. \quad (19)$$

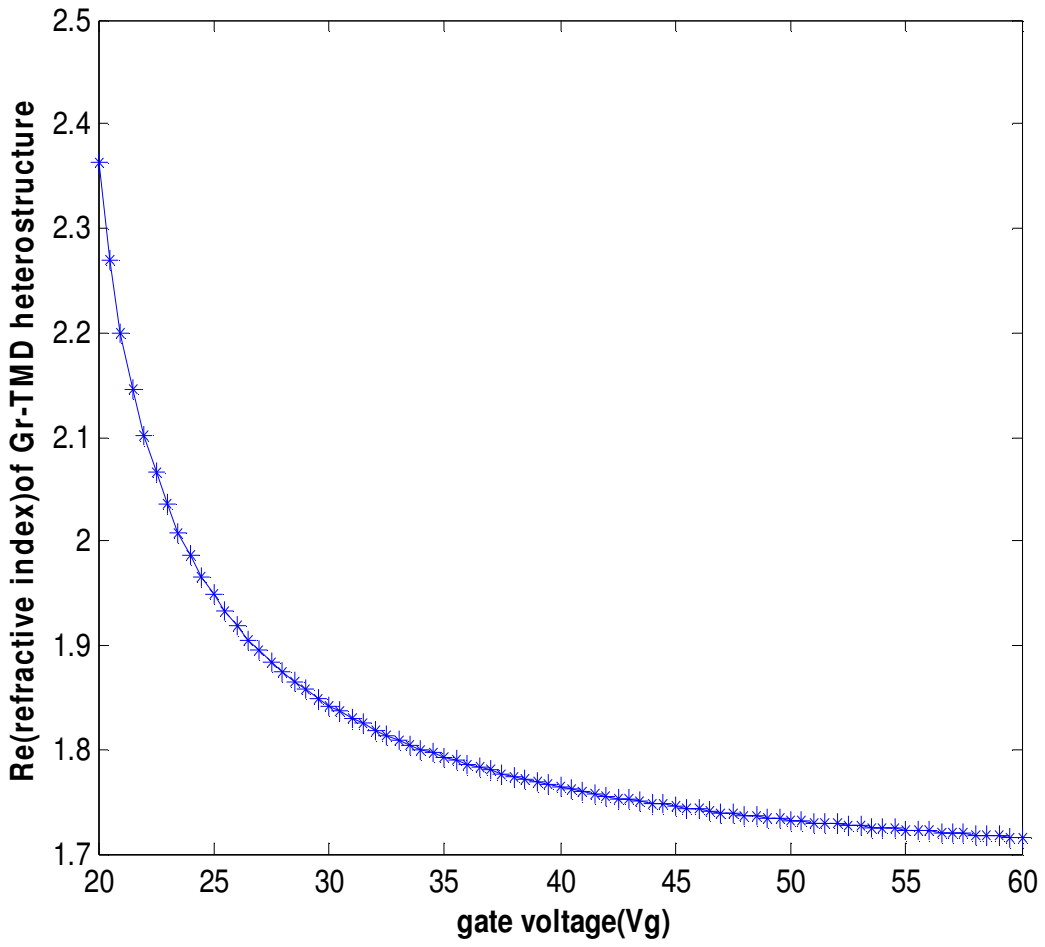

**Figure 6**. A 2D plot of the real part of the refractive index of Gr-TMD heterostructures as a function of the gate voltage.

The electro-optic modulators exploit electro-optic effects to electrically control the light properties. They are particularly desirable for data communication link applications. Thus far, 2D material based electro-optic modulators have been demonstrated mainly by utilizing the gate-tunable electro-absorption effect in graphene **[23,24]**. Recently, the tunability of the refractive index (RI)of graphene has been experimentally demonstrated by gating**[25,26]**. Other electro-optic effects in 2D materials, such as Franz-Keldysh effect and quantum-confined Stark effect**[27]** , are also possible for light modulation, but have not yet been experimentally demonstrated. Our graph in Figure 6 shows the tunability of refractive index of the graphene based heterostructure, which is the most important parameter for the design of photonic devices for optical communications. It also indicates the possibility of using (tunable) Gr-TMD material for the electro-refractive phase modulation.

Yet another potential application of the graphene-transition metal dichalcogenide (TMDC)-graphene hybrid nanostructures is the plasmonic meta-surface for ultra-sensitive bio-sensing. The sensing principle is the utilization of the exponentially decaying fields of a surface plasmon wave (SPW)propagating along interface, which is highly sensitive to the ambient refractive index variations, such as induced by bio affinity interactions at the sensor surface. An important sensor performance parameter is the reciprocal of full width at half maximum (FWHM). Therefore, for a SPW sensor with excellent performance the FWHM should be as small as possible. In I, we have calculated the spectral function. Close to the Plasmon frequency $\omega'_{m,pl}$ , we obtained the expression for the spectral function

$$\aleph(\omega',\mu,T) \sim \frac{(H(\mu,\omega',T))}{[(\omega'_{m,pl}-\omega')^2\left(\frac{C^{-8/3}(a|q|)^{\frac{2}{3}}}{(Q(\mu,T,M))^{\frac{2}{3}}}\right)+(H(\mu,\omega',T))^2]} , \qquad (20)$$

which resembles a Lorentzian with the FWHM and height, respectively, given by the functions $\Gamma(\mu,\omega',T)$and $(H(\mu,\omega',T))^{-1}$ where

$$\Gamma(\mu,\omega',T) \sim \mathcal{C}^{4/3}\left(\frac{(Q(\mu,T,M))^{\frac{1}{3}}}{\hbar\omega'\kappa\, v_F\, \varepsilon_r(a|\boldsymbol{q}|)^{\frac{1}{3}}}\right)(2\pi\alpha c) \sim \left(\frac{(Q(\mu,T,M))^{\frac{1}{3}}}{(\hbar\omega')^{\frac{3}{2}}\kappa\, v_F\, \varepsilon_r}\right)(2\pi\alpha c)\mathcal{C}^{4/3},\ \mathcal{C} = \pi g_v\left(\frac{\frac{e^2}{2\,a\varepsilon_0\varepsilon_r}}{\left(\frac{\hbar v_F}{a}\right)}\right). \quad (20)$$

$$Q(\mu,T=0,M) = \left\{\sum_s \frac{\left[\left(\mu'-s\sqrt{\left(\frac{z_0(M)}{2}\right)}\lambda_R\right)^2 - \lambda_{-s}(\xi=1,M)^2\right]^{\frac{3}{2}}}{\lambda_{-s}(\xi,M)^3}\right\},\ \mu' = \mu/\hbar v_F a^{-1}.$$

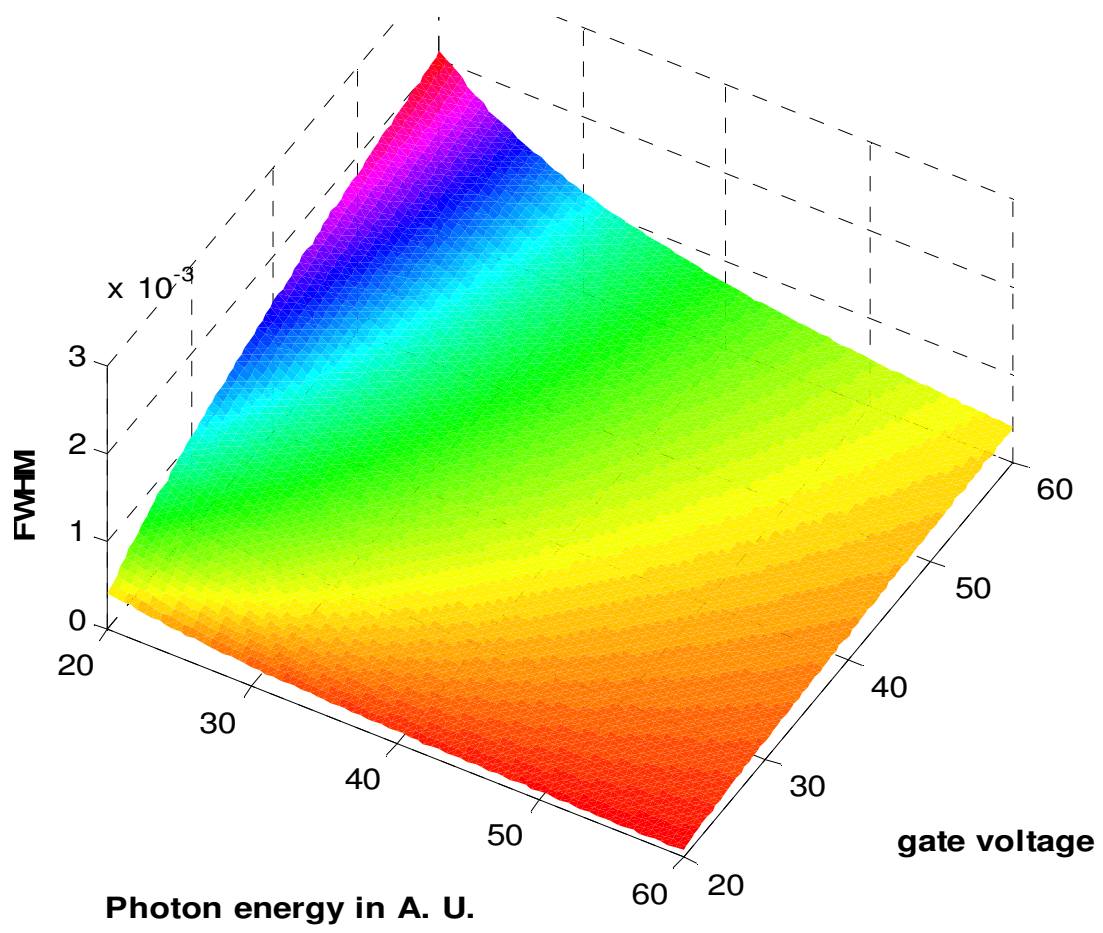

**Figure 7**. A 3 D plot of the FWHM as a function of the photon energy and the gate voltage.

The spectral function $\aleph(\omega',\mu,T,q)$ is symmetrical about the position of its maximum. This function characterizes the probability of electrons to undergo surface excitation in surface region. We note that FWHM could be controlled by changing the chemical potential through the electrostatic / chemical doping and could be made as narrow as is required. Furthermore, we find that the full width at half maximum is narrow at high photon energy and low gate voltage (see Figure 7). It may now be noted that though metal-dielectric based SPR has been widely employed for sensing applications **[28,29]**, such as gas sensing, temperature sensing, and bio-sensing, during the last two decades due to its high sensitivity and reliability, a SPW sensor based on the Gr-TMD heterostructure, however, has greater advantage as it is likely to have good (tunable) performance.

## 6. Conclusion

It has been also demonstrated ( in I and ref. **[8]**) that the exchange field can be used for efficient tuning of the band gap, the Thomas-Fermi screening length and the intra-band plasmon frequency. The intra-band conductivity and absorbance, too, can be controlled by the exchange field. A direct, functional electric field control of magnetism at the nano-scale is needed for the effective demonstration of these results related to the exchange-field dependence. The magnetic multi-ferroics, like $BiFeO_3$ (BFO) have piqued the interest of the researchers world-wide with the promise of the coupling between the magnetic and electric order parameters.

In conclusion, the Gr-TMD plasmons (optical as well as THz varieties) have unusual properties and offer promising prospects for plasmonic applications covering a wide frequency range. It is useful for manipulating electromagnetic signals at the deep-subwavelength scale owing to its remarkable physical properties, e.g., high carrier mobility and electrostatically tunable optical properties. Thus, the extraordinary properties of SPs in Gr-TMD, plus its good flexibility, and stability make it a good candidate for varieties of applications, including THz technology, energy storage, biotechnology, medical sciences, electronics, optics, and so on.